\documentclass[preprint2]{aastex62}

\usepackage{graphicx}
\usepackage{bm}
\usepackage{amssymb}
\usepackage{color}
\usepackage{natbib}
\usepackage{amsmath}
\usepackage{enumitem}   

\graphicspath{{./}{}}

\received{\today}

\shortauthors{Franci et al.}

\begin{document}

\title{Modeling Kelvin-Helmholtz instability-driven turbulence with hybrid simulations of Alfv\'enic turbulence}

\correspondingauthor{Luca Franci}
\email{l.franci@qmul.ac.uk}

\author[0000-0002-0786-7307]{Luca Franci}
\affiliation{School of Physics and Astronomy, Queen Mary University of London, London, UK}
\affiliation{INAF, Osservatorio Astrofisico di Arcetri, Firenze, Italy}

\author[0000-0002-5702-5802]{Julia E. Stawarz}
\affiliation{Department of Physics, Imperial College London, London, UK}

\author[0000-0002-7969-7415]{Emanuele Papini}
\affiliation{Dipartimento di Fisica e Astronomia, Universit\`a degli Studi di Firenze, Sesto Fiorentino, Italy}

\author[0000-0002-5608-0834]{Petr Hellinger}
\affiliation{Institute of Atmospheric Physics, The Czech Academy of Sciences, Prague, Czech Republic}
\affiliation{Astronomical Institute,The Czech Academy of Sciences, Prague, Czech Republic}

\author[0000-0003-4550-2947]{Takuma Nakamura}
\affiliation{Space Research Institute, Austrian Academy of Sciences, Graz, Austria}

\author[0000-0002-8175-9056]{David Burgess}
\affiliation{School of Physics and Astronomy, Queen Mary University of London, London, UK} 

\author[0000-0002-1322-8712]{Simone Landi}
\affiliation{Dipartimento di Fisica e Astronomia, Universit\`a degli Studi di Firenze, Sesto Fiorentino, Italy}
\affiliation{INAF, Osservatorio Astrofisico di Arcetri, Firenze, Italy}

\author[0000-0003-4380-4837]{Andrea Verdini}
\affiliation{Dipartimento di Fisica e Astronomia, Universit\`a degli Studi di Firenze, Sesto Fiorentino, Italy}

\author[0000-0002-1322-8712]{Lorenzo Matteini}
\affiliation{Department of Physics, Imperial College London, London, UK}

\author[0000-0002-3096-8579]{Robert Ergun}
\affiliation{Laboratory of Atmospheric and Space Sciences, University of Colorado Boulder, Boulder, Colorado, USA}

\author[0000-0003-2713-7966]{Olivier Le Contel}
\affiliation{Laboratoire de Physique des Plasmas, CNRS/Ecole Polytechnique/Sorbonne Université/Univ. Paris-Sud/Observatoire de Paris, Paris, France}

\author[0000-0001-5617-9765]{Per-Arne Lindqvist}
\affiliation{KTH Royal Institute of Technology, SE-100 44 Stockholm, Sweden}

\begin{abstract}
Magnetospheric Multiscale (MMS) observations of plasma turbulence generated by a Kelvin-Helmholtz (KH) event at the Earth's magnetopause are compared with a high-resolution two-dimensional (2D) hybrid direct numerical simulation (DNS) of decaying plasma turbulence driven by large-scale balanced Alfv\'enic fluctuations. The simulation, set up with four observation-driven physical parameters (ion and electron betas, turbulence strength, and injection scale) exhibits a quantitative agreement on the spectral, intermittency, and cascade-rate properties with in situ observations, despite the different driving mechanisms. Such agreement demonstrates a certain universality of the turbulent cascade from magnetohydrodynamic (MHD) to sub-ion scales, whose properties are mainly determined by the selected parameters, also indicating that the KH instability-driven turbulence has a quasi-2D nature. The validity of the Taylor hypothesis in the sub-ion spatial range suggests that the fluctuations at sub-ion scales have predominantly low frequencies, consistent with a kinetic Alfv\'en wave-like nature or with quasi-static structures. Finally, the third-order structure function analysis indicates that the cascade rate of the turbulence generated by a KH event in the magnetopause is an order of magnitude larger than in the ambient magnetosheath.
\end{abstract}


\section{Introduction}
\label{sec:intro}

Plasma turbulence is a fundamental phenomenon in many astrophysical systems, including the solar wind \citep[e.g.][and references therein]{Matthaeus_Velli_2011,Bruno_Carbone_2013,Kiyani_al_2015,Chen_2016,Verscharen_al_2019} and the Earth's magnetosphere \citep[e.g.,][]{Voros_al_2006,Zimbardo_al_2010,Breuillard_al_2018,Macek_al_2018,Pollock_al_2018,Roberts_al_2018}, where it can be investigated by in-situ spacecraft observations.
A process responsible for the generation of turbulence is the Kelvin-Helmoholtz (KH) instability, which in the Earth's magnetopause is driven by the velocity shear between the shocked solar wind and the magnetosphere \citep[e.g.][and references therein]{Nakamura_al_2019}.

Plasma turbulence has been studied for decades via theoretical
modeling by direct numerical simulations (DNS), with  different methods in different regimes and ranges of scales~\citep[e.g.][]{Gary_al_2012,Boldyrev_al_2013,Karimabadi_al_2013,TenBarge_Howes_2013,Vasconez_al_2014,Vasquez_al_2014,Franci_al_2015b,Franci_al_2015a,Parashar_al_2015,Passot_al_2015,Servidio_al_2015,Told_al_2015,Wan_al_2015,Matthaeus_al_2016,Cerri_al_2017a,Kobayashi_al_2017,Pucci_al_2017,Valentini_al_2017,Groselj_al_2018,Passot_al_2018,Arzamasskiy_al_2019,Papini_al_2019,Roytershteyn_al_2019,Zhdankin_al_2019}. 
Both observations and DNS deliver
spectra of the solar wind plasma and electromagnetic fluctuations exhibiting
clear power laws over several decades in frequency (or,
correspondingly, in wavenumber), with a transition (break) at the proton characteristic
scales \citep[e.g.,][]{Alexandrova_al_2009,Sahraoui_al_2010,Chen_2016}. Such power-law behavior suggests that a turbulent cascade is at play at large fluid scales and
continues all the way down to particle characteristic scales, where kinetic effects become important.  
Despite a certain variability, mainly due to different plasma conditions, a general consensus has been achieved on the slope of different fields at scales larger and smaller than the ion-scale break, i.e., in the MHD/inertial and in the sub-ion range, respectively 
\citep[e.g.,][]{Bale_al_2005,Podesta_al_2007,Alexandrova_al_2009,Chandran_al_2009,Sahraoui_al_2010,Boldyrev_al_2011,Chen_al_2012,Chen_al_2013,Safrankova_al_2015,Chen_2016,Safrankova_al_2016,Chen_al_2017,Matteini_al_2017}.

The cascade is characterized by intermittency~\citep[e.g.][]{Matthaeus_al_2015}, expressed by non-Gaussian probability distribution functions (PDFs) of the turbulent fluctuations, due to the presence of discontinuities and current sheets at different scales. Such departure from a normal distribution is observed to increase as smaller scales are approached, likely due to the presence of ion-scale current sheets and magnetic reconnection sites \citep[e.g.][and rererences therein]{Bruno_2019}.

Theoretical predictions for the turbulent cascade rate can be obtained by
assuming incompressibility, homogeneity, and isotropy. These derive from a law that links the cascade/dissipation rate
to 3rd-order mixed structure functions involving magnetic and velocity fluctuations
\citep{Karman_Howarth_1938,Politano_Pouquet_1998}.
Such relation is observed
as a linear scaling of the structure functions
with the separation scale in the inertial range \citep{MacBride_al_2005,SorrisoValvo_al_2007,Verdini_al_2015}, 
where the cascade rate represents the coefficient (apart for a factor $-4/3$).


Although spacecraft observations represent unique opportunities  
to measure in-situ plasma properties, they provide only
single-point or, at most, few-point measurements, performed at specific moments in time. Often, they are not equipped to measure all fields simultaneously and/or they do so with different time resolutions. DNS 
provide fundamental complementary information, e.g., particle distribution functions and their moments and
electromagnetic fields at many millions of grid points simultaneously, 
with the same spatial and time resolution, 
accompanied by two-dimensional (2D) and/or three-dimensional (3D) images.
Thanks to recent increased accuracy in both spacecraft
measurements and DNS, we can now fully exploit their synergy to probe the plasma kinetic scales by interpreting observations with numerical modelling.

Here, we compare a high-resolution 2D hybrid (particle-in-cell protons, mass-less fluid electrons) DNS of decaying Alfv\'enic turbulence~\citep[cf.][]{Franci_al_2015b} with MMS observations of plasma turbulence generated by a KH event in the Earth's magnetopause~\citep{Stawarz_al_2016}. Unlike in observations and previous DNS of KH-driven turbulence~\citep[e.g.,][]{Nakamura_al_2011,Nakamura_al_2013}, here the large-scale energy injection is not due to a super-Alfv\'enic vortex flow. We investigate numerically the development of the turbulent cascade after the initial injection has occurred by setting the ion and electron plasma betas, the turbulence strength, and the injection scale to their observational values, to mimic the observed plasma conditions.

The paper is organized as follows. In Section \ref{sec:data}, we describe how the observational and numerical datasets have been selected and post-processed. In Section \ref{sec:results}, we compare the MMS and DNS results, with particular focus on the power spectra of electromagnetic and plasma fluctuations (Subsection \ref{subsec:spectra}), intermittency (\ref{subsec:intermittency}), and cascade rate (\ref{subsec:yaglom}). Finally, in Section \ref{sec:discussion}, we discuss and summarize our findings and their physical implications.

\section{Observational and numerical datasets}
\label{sec:data}

The observational dataset consists of fifty-four subintervals of
high-resolution ``burst'' data collected by MMS on the duskside magnetopause on 8 September 2015. It contains magnetic field measurements from the fluxgate \citep{Russell_al_2016} and the searchcoil \citep{LeContel_al_2016} magnetometers at $1/128$ and $1/8192$ s cadences respectively, electric field measurements from the electric
field double probes \citep{Ergun_al_2016, Lindqvist_al_2016} at
$1/8192$ s cadence, and ion and electron particle moments, as measured
by the fast plasma investigation \citep{Pollock_al_2016}, at $0.15$ and
$0.03$ s cadences respectively. Subintervals were manually selected from
a $\sim1.5$ hour-long interval of burst data, in which a continuous train of
Kelvin-Helmholtz (KH) waves were observed. The subintervals are the
same as those examined by \citet{Stawarz_al_2016} and were selected so
as to avoid periodic compressed current sheets associated with the
large-scale KH wave, which would skew the statistics if included in
the analysis. Aspects of this KH event, including magnetic
reconnection \citep{Eriksson_al_2016a, Eriksson_al_2016b, Li_al_2016, Vernisse_al_2016, Sturner_al_2018}, turbulence \citep{Stawarz_al_2016,SorrisoValvo_al_2019},
waves \citep{Wilder_al_2016}, and mass transport \citep{Nakamura_al_2017a,Nakamura_al_2017b} 
properties, have been examined in a number of previous
studies. \citet{Eriksson_al_2016a} provide an overview of the properties
of the overall KH instability within this event, which is characterized by an average ion inertial length of $d_i \sim 65$ km and a KH wavelength of the order of $300\,d_i$.

\begin{figure}[ht!]
\begin{center}
\includegraphics[trim={0.5cm 0.5cm 0.5cm 0.5cm},clip,width=\linewidth]{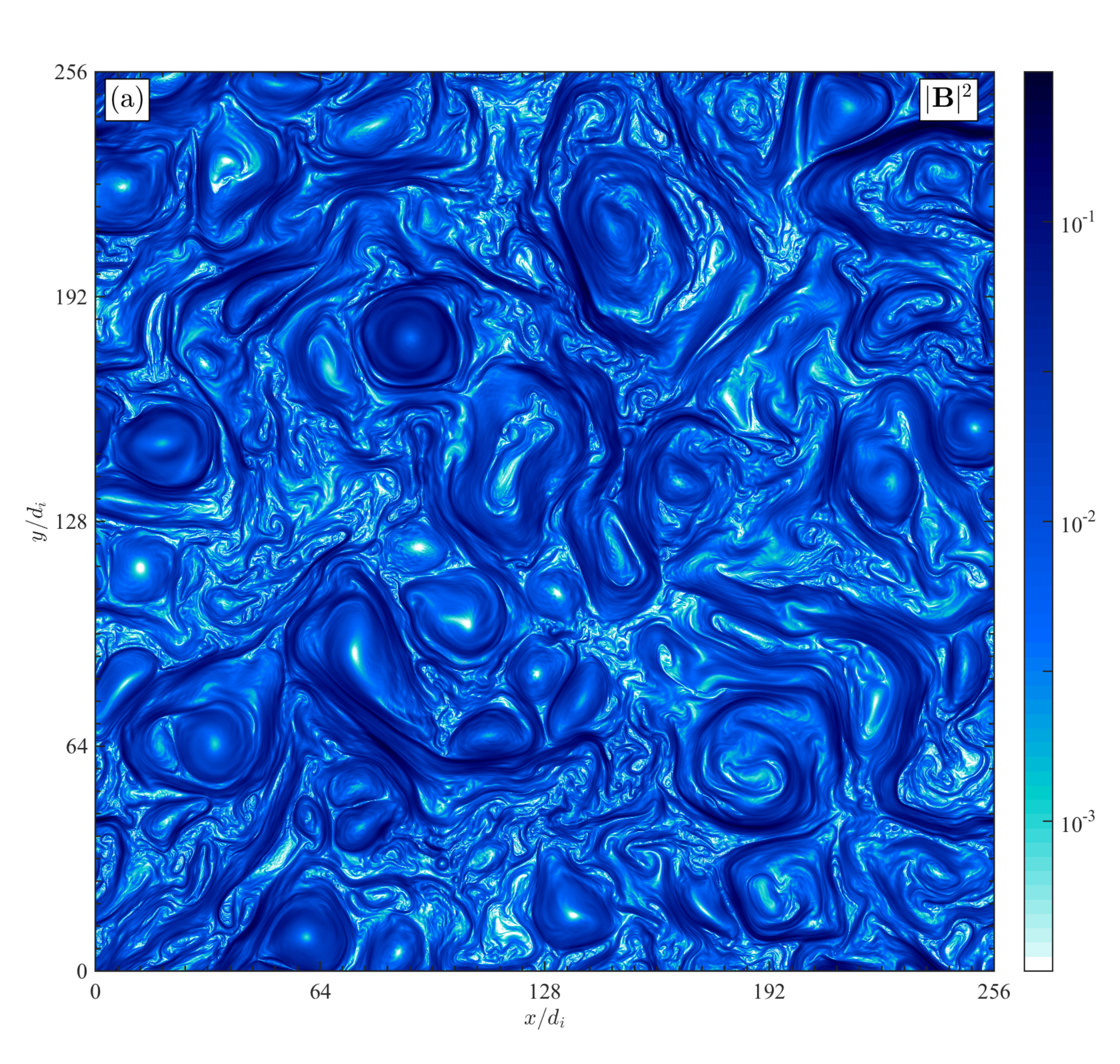}\\
\includegraphics[trim={0.5cm 0.5cm 0.5cm 0.5cm},clip,width=\linewidth]{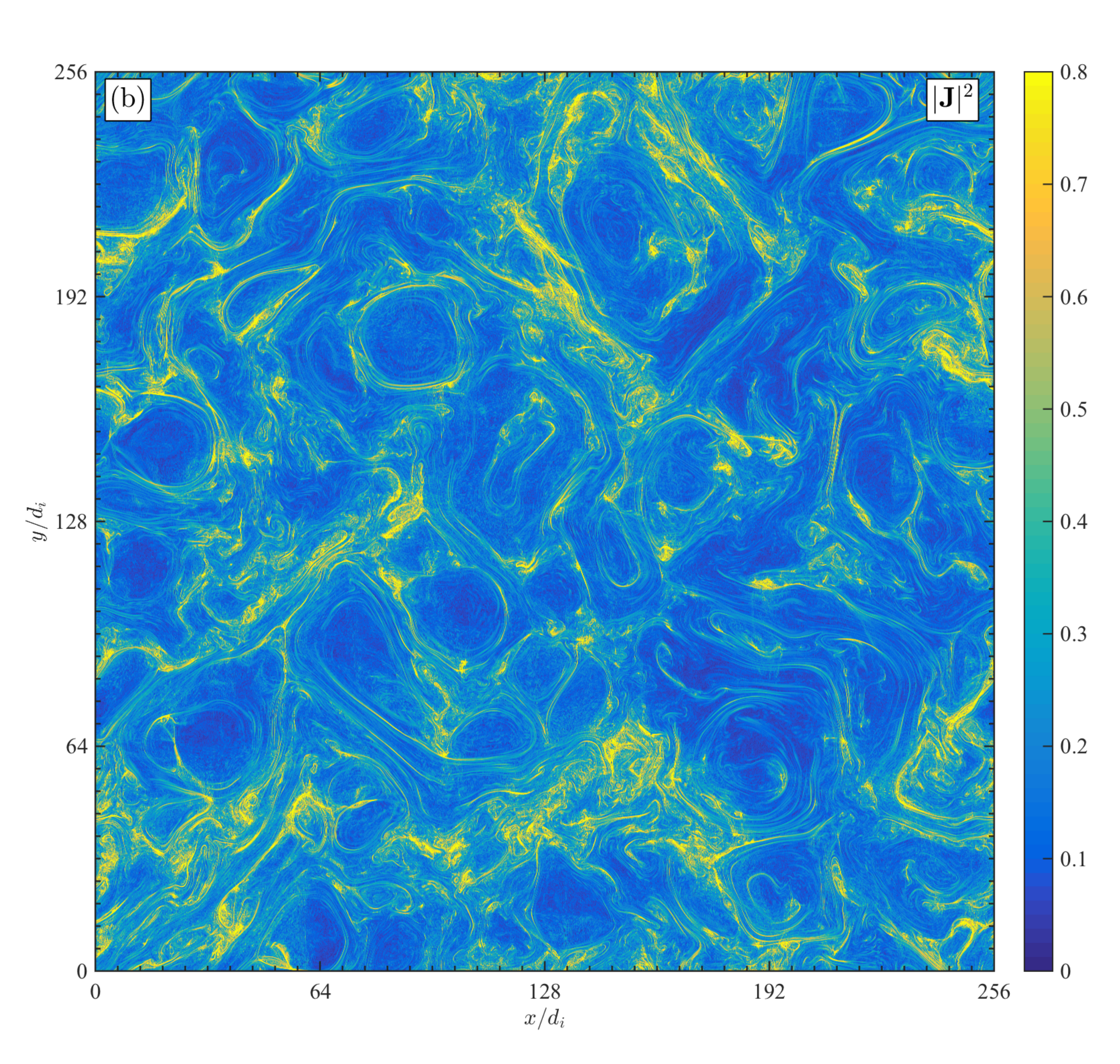}
\end{center}
\caption{Contour plots of the magnitude of the magnetic fluctuations (a) and of the current density (b) in the whole DNS domain at the final time.}
\label{fig:realspace}
\end{figure}

The numerical dataset was obtained from a high-resolution 
2D DNS of plasma turbulence 
performed with the hybrid (particle-in-cell ions and mass-less fluid electrons)
code CAMELIA \citep{Franci_al_2018b}.
The temporal and spatial units are the
inverse ion gyrofrequency, $\Omega_i^{-1}$, and the 
ion inertial length, $d_i$.  The initial conditions 
and the normalization units for the different fields are the same as in
~\citet{Franci_al_2015b},
i.e., a homogeneous plasma embedded in a uniform out-of-plane ambient
magnetic field, $\bm{B}_0$, perturbed by balanced Alfv\'enic fluctuations.   
The numerical parameters are: $16384$ particle-per-cell (ppc), $4096\times4096$ grid points, a box size of $256\,d_i$ (of the same order of the KH wavelength for the observed event, see above), a spatial resolution
of $d_i / 16$, a time resolution of $\Delta t=0.01 \, \Omega_i^{-1}$ for the
particle advance and $\Delta t_B = \Delta t/10$ for the magnetic field advance, and a resistivity $\eta=2 \times 10^{-4}~4\pi v_Ac^{-1}\Omega_i^{-1}$, where $v_A$ is the Alfv\'en velocity.  
The following physical parameters have been set accordingly to their
observational values: ion beta
$\beta_i=0.42$, electron beta $\beta_{\rm
  e}=0.065$, amplitude of the initial Alfv\'enic fluctuations
$\bm{B}^{\mathrm{rms}}/\bm{B}_0 \sim 0.14$, and injection scale $k_\perp d_i \lesssim 0.3$. 
The development of the turbulent cascade is qualitatively the same as
in \citet{Franci_al_2017}. The simulation has been run until $350 \; \Omega_i^{-1}$, when the root mean
square value of the current density has reached a
plateau. At this time, a turbulent quasy-stationary state has been achieved, characterized by magnetic field structures at all scales (Fig.~\ref{fig:realspace}(a)) and small scale current structures between vortices (Fig.~\ref{fig:realspace}(b)). We perform our analysis at $300 \; \Omega_i^{-1}$, without performing any time average. We have verified that spectra and other properties do not change significantly between that time and
the end of the simulation.

\section{Results}
\label{sec:results}
\subsection{Spectral properties}
\label{subsec:spectra}

\begin{figure}[ht!]
\begin{center}
\includegraphics[width=\linewidth]{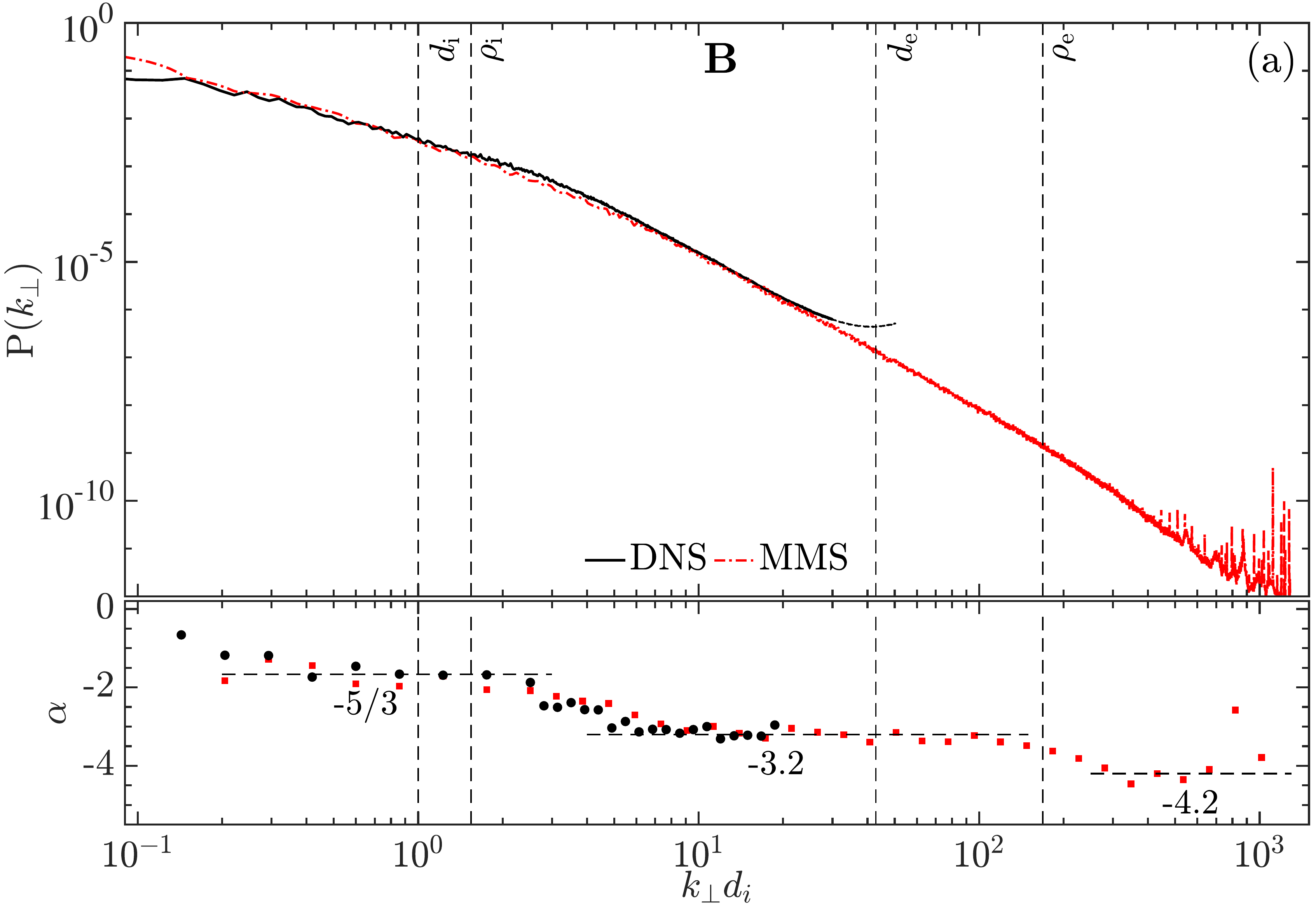}\\
\includegraphics[width=\linewidth]{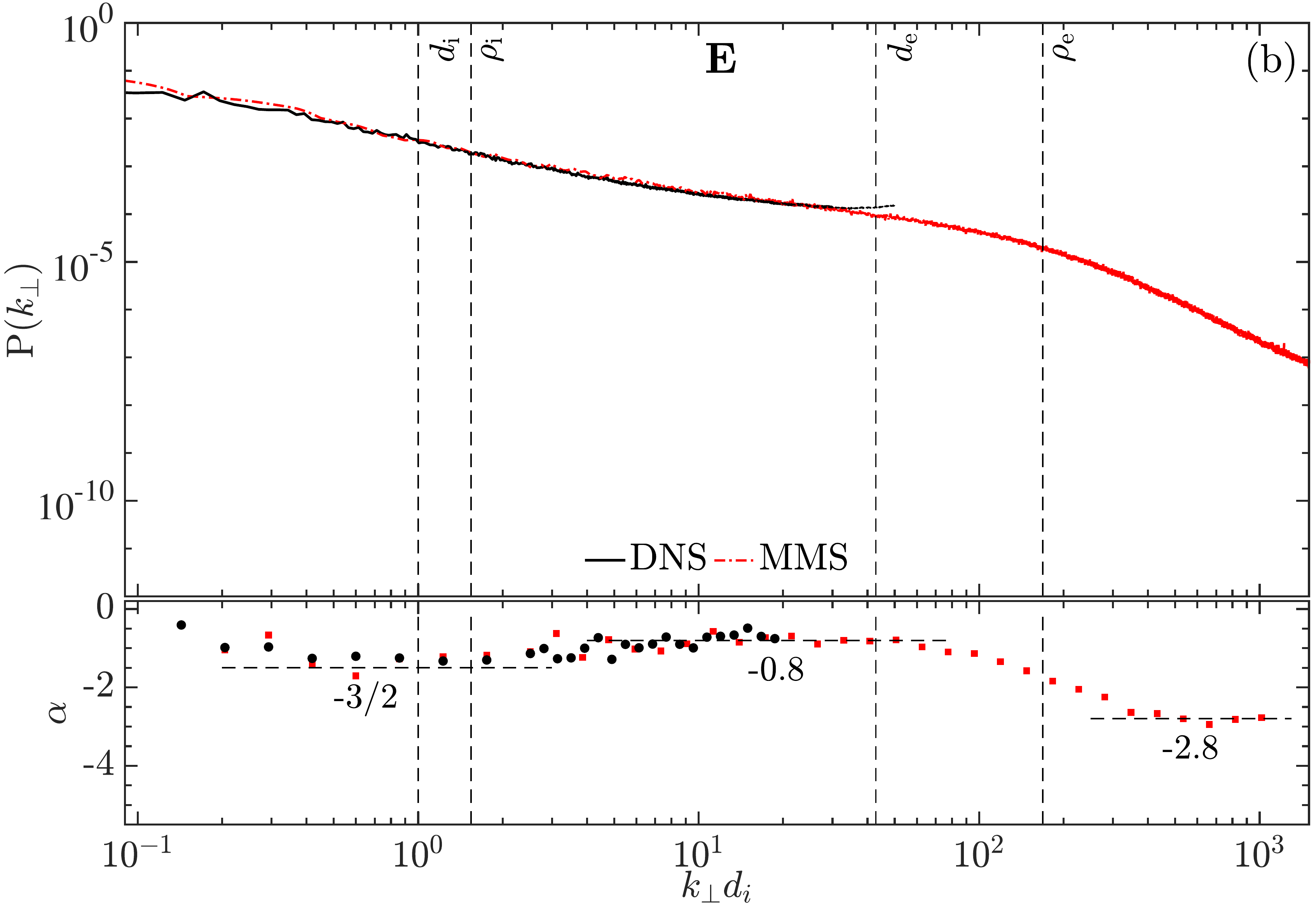}
\end{center}
\caption{Spectral properties of the electromagnetic fluctuations: magnetic field (a) and electric field (b). Upper panels: 1D MMS and DNS spectra. Bottom panels: local slope, $\alpha$.}
\label{fig:spectra_em}
\end{figure}

\begin{figure}[ht!]
\begin{center}
\includegraphics[width=\linewidth]{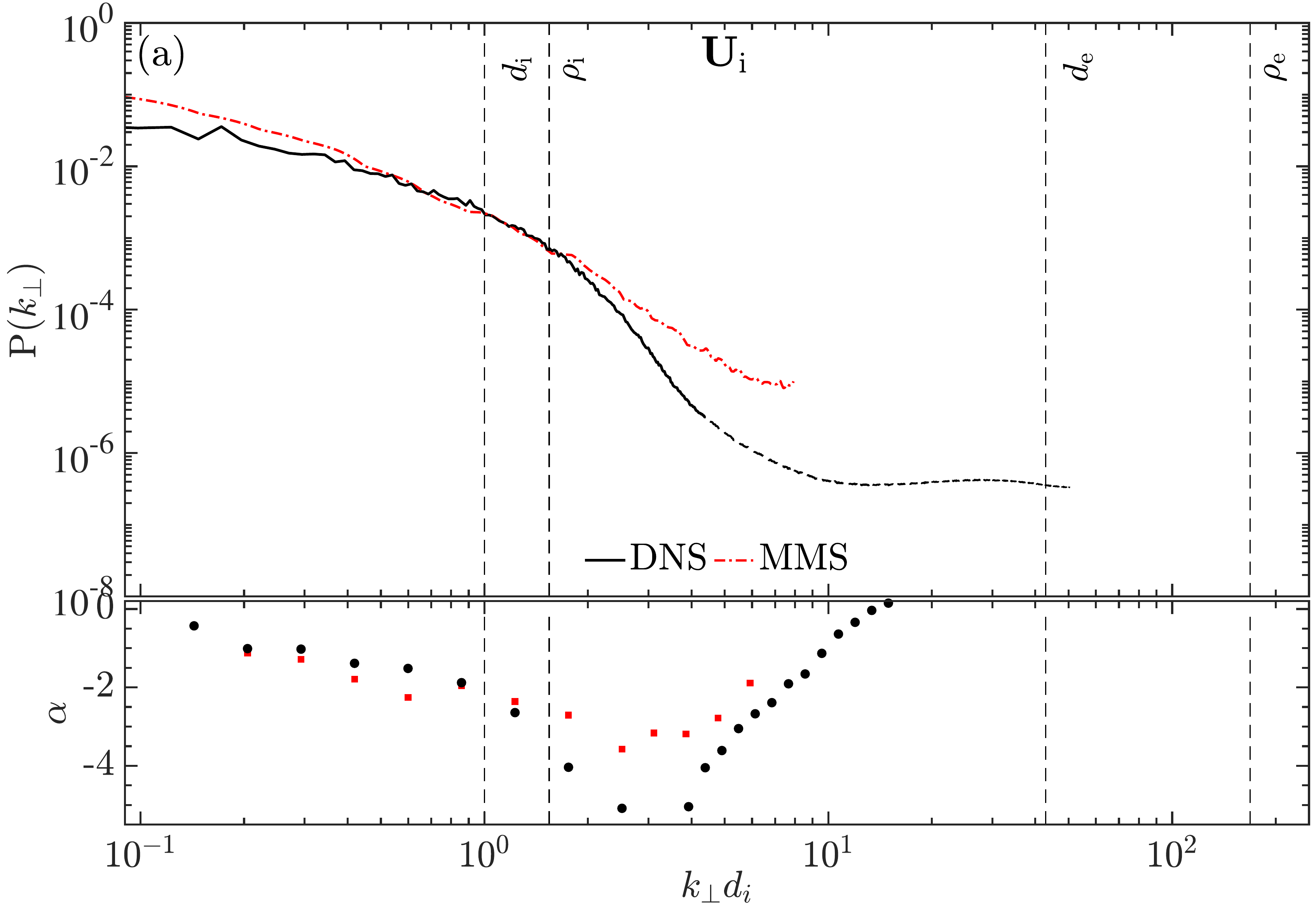}\\
\includegraphics[width=\linewidth]{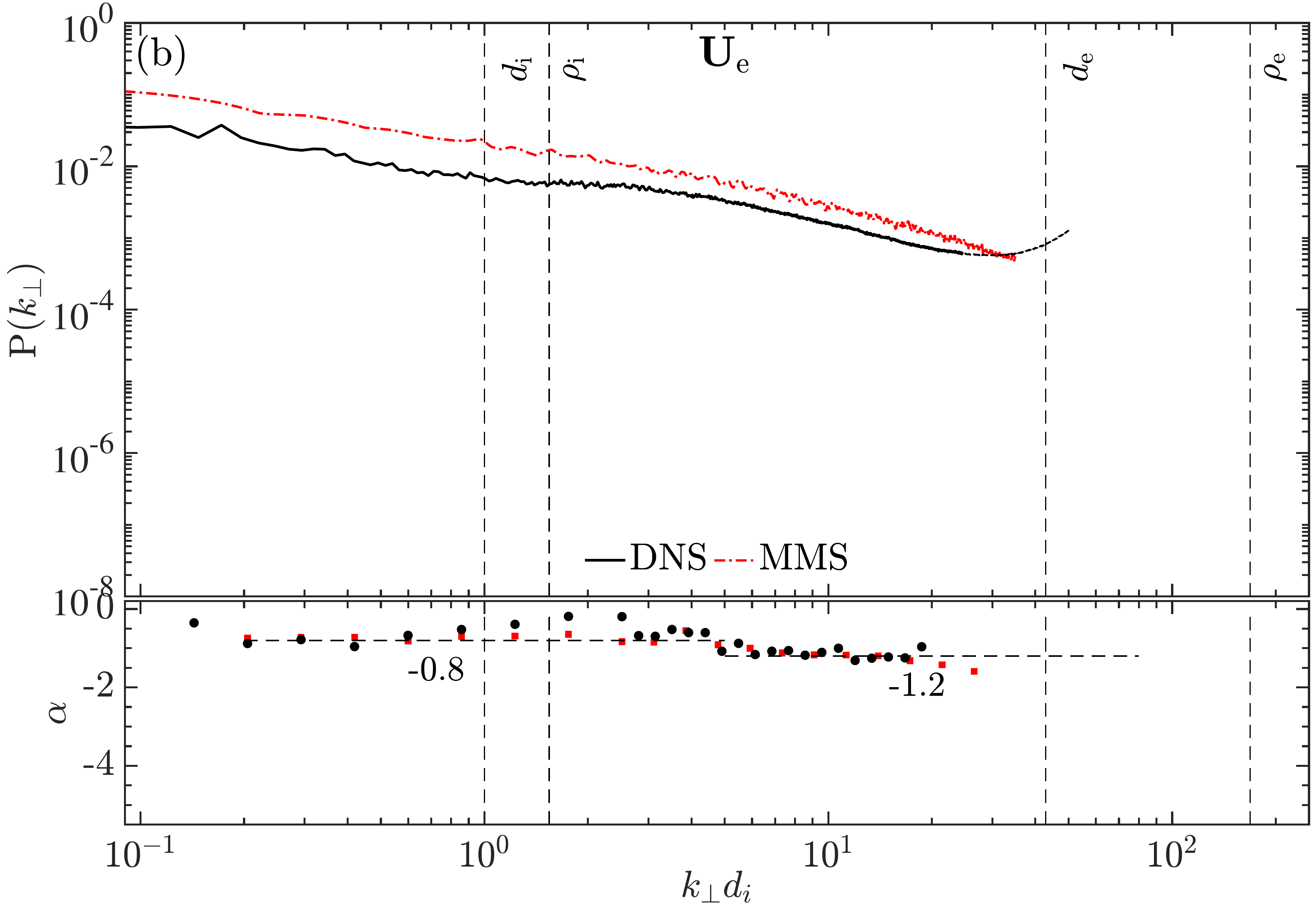}\\
\includegraphics[width=\linewidth]{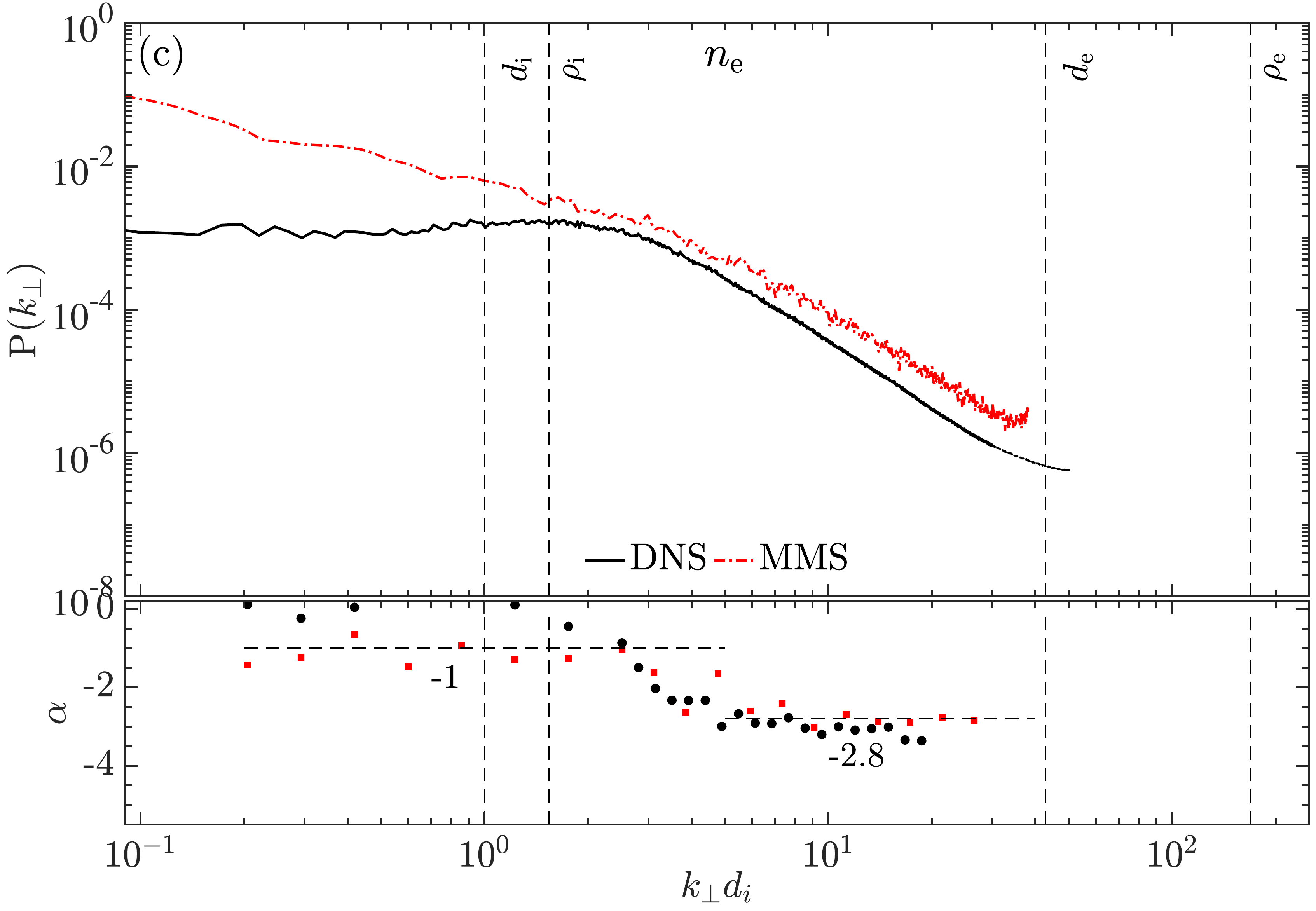}
\end{center}
\caption{The same as in Fig.\ref{fig:spectra_em} (but with different vertical and horizontal scales), for the plasma fluctuations: ion (a) and  electron velocity (b), and electron density (c).}
\label{fig:spectra_plasma}
\end{figure}

The observational
and numerical spectral properties of the electromagnetic and plasma fluctuations are shown in Fig.~\ref{fig:spectra_em} and Fig.~\ref{fig:spectra_plasma}, respectively, using red dash-dotted lines for MMS and black solid ones for the DNS. Four vertical dashed lines mark the particle characteristic scales, i.e., from small to large wavenumbers, the ion inertial length, $d_i$, the ion gyroradius, $\rho_i = \sqrt{\beta_i} d_i$, the electron inertial length, $d_e$, and the electron gyroradius, $\rho_e = \sqrt{\beta_e} d_e$.
In the top panels, the 1D power spectra of each field are compared. The MMS spectra have been converted from frequency, $f$, to perpendicular
 wavenumber with respect to the ambient magnetic field, $k_\perp$, using the ``frozen flow'' Taylor hypothesis \citep{Taylor_al_1938}. Under the assumptions that the average perpendicular ion bulk flow velocity, $U_{0,\perp} = 170
\,\mathrm{km/s}$, is much larger than its parallel counterpart \citep[e.g.][]{Stawarz_al_2016} and that $k_\perp \gg k_\parallel$ (consistent with Alfv\'enic turbulence), such hyphotesis yields $k_\perp = 2 \pi f/ U_{0,\perp}$. The DNS spectra have not been rescaled in amplitude and their portions affected by numerical noise are drawn as dotted lines.
In the bottom panels we compare the values of the local spectral index, $\alpha$, obtained by performing many power-law fits on small intervals in $k_\perp$.
For each field, horizontal power laws with characteristic slopes are drawn as a reference.

In Fig.~\ref{fig:spectra_em}(a),
we analyze the fluctuations of the magnetic field, $\bm{B}$.  
The MMS spectrum exhibits
a triple power-law-like behavior over four
decades in wavenumber, with a first steepening around
the ion scales and a
second one around the electron scales.  It is closely followed by the DNS spectrum over the two full decades
where the numerical noise is negligible, i.e., in the wavenumber interval $0.3 \lesssim k_\perp
d_i \lesssim 30$. 
The bottom panel shows that the first power-law slope is compatible with the $-5/3$ prediction for the intertial-range turbulent cascade \citep{Goldreich_Sridhar_1995}. 
A transition is observed in
correspondence of $k_\perp d_i \sim 3$, consistently with previous DNS with a similar plasma beta \citep{Franci_al_2015b, Franci_al_2016b}.
A second, steeper ($\alpha \sim -3.2$), power law is observed at sub-ion scales, 
extended for over a decade and a
half in MMS, up to $k_\perp d_i \sim 100$, and over slightly less than a decade in the DNS, until the resolution limit is reached. A second transition
is observed in the data just before the electron scales are met, followed by a third, much narrower interval where the instrumental noise becomes important, 
so that no clear power law can be inferred, despite a hint of a further
steepening. 

Three distinct, clearer, power-law intervals are 
observed in the MMS spectrum of the electric field, $\bm{E}$ 
(Fig.~\ref{fig:spectra_em}(b)). The agreement with the DNS spectrum is particularly
remarkable, since in the hybrid model $\bm{E}$ is computed from the generalized Ohm's law involving the other fields and their derivatives, so it's more sensitive to both physical parameters (e.g., ion and electron beta) and small-scale numerical noise (e.g., from the density and the current density). The first interval is compatible with a power law with
slope $-3/2$, followed by a flattening to
$\sim -0.8$ at sub-ion scales.  Both these spectral indices are consistent with theoretical predictions
from the generalized Ohm's law \citep{Franci_al_2015b}.  In correspondence with 
the electron scales, the MMS spectrum shows a second transition and a
steepening, hinting at a third power-law interval with slope $\sim-2.8$, although less
extended than a decade, for $300 \lesssim k_\perp d_i \lesssim 1000$.

The spectra of the ion bulk velocity, $\bm{U}_i$ (Fig.~\ref{fig:spectra_plasma}(a)), show no clear extended power-law interval, neither in the inertial range nor in the kinetic one. The level of the ion bulk velocity fluctuations strongly drops at $k_\perp d_i \sim 1$, so there is not even a full decade separating the break from the injection scale. Approaching the kinetic scales, both the MMS and the DNS spectra flatten, due to instrumental and numerical noise, respectively. In the range of scales not affected by noise, the two spectra are still in quite good qualitative agreement. 

Figure \ref{fig:spectra_plasma}(b) compares the MMS and DNS spectra of the electron bulk velocity, which in the hybrid model is computed as $\bm{U}_e = \bm{U}_i - (\nabla \times \bm{B}) / n$. By chance, the maximum wavenumber is the same, since the spatial resolution of the DNS corresponds to the observational time resolution for this field. A small discrepancy in the level of fluctuations is observed over the whole range of scales, whose origin is not clear. The DNS spectrum of $\bm{U}_e$ in the kinetic range is just that of $\bm{B}$ multiplied by $k_\perp^2$, consistently with the fact that the ion bulk velocity fluctuations are negligible at those scales and so the current is almost entirely supported by the electron bulk motion, i.e., $\nabla \times \bm{B} = \bm{J} = n_i \bm{U}_i - n_e \bm{U}_e \sim - n \,\bm{U}_e$. Despite the slightly different level, the slope of the two spectra of $\bm{U}_e$ is almost exactly the same at all scales, with just a small bump around the ion scales in the DNS. 

Finally, the spectral properties of the density fluctuations are compared in Fig.~\ref{fig:spectra_plasma}(c). MMS measures both the ion, $n_{\rm{i}}$, and the electron density fluctuations, $n_{\rm{e}}$, with two different instruments and resolutions. In the hybrid model, since the electrons are treated as a mass-less fluid, these two quantities are assumed to be the same. Therefore, here we only show the observed electron density, which has a higher time resolution, corresponding to the DNS spatial resolution.
%
\begin{figure*}[ht!]
\begin{center}
\includegraphics[trim={0 1cm 0 0},clip,width=0.4\linewidth]{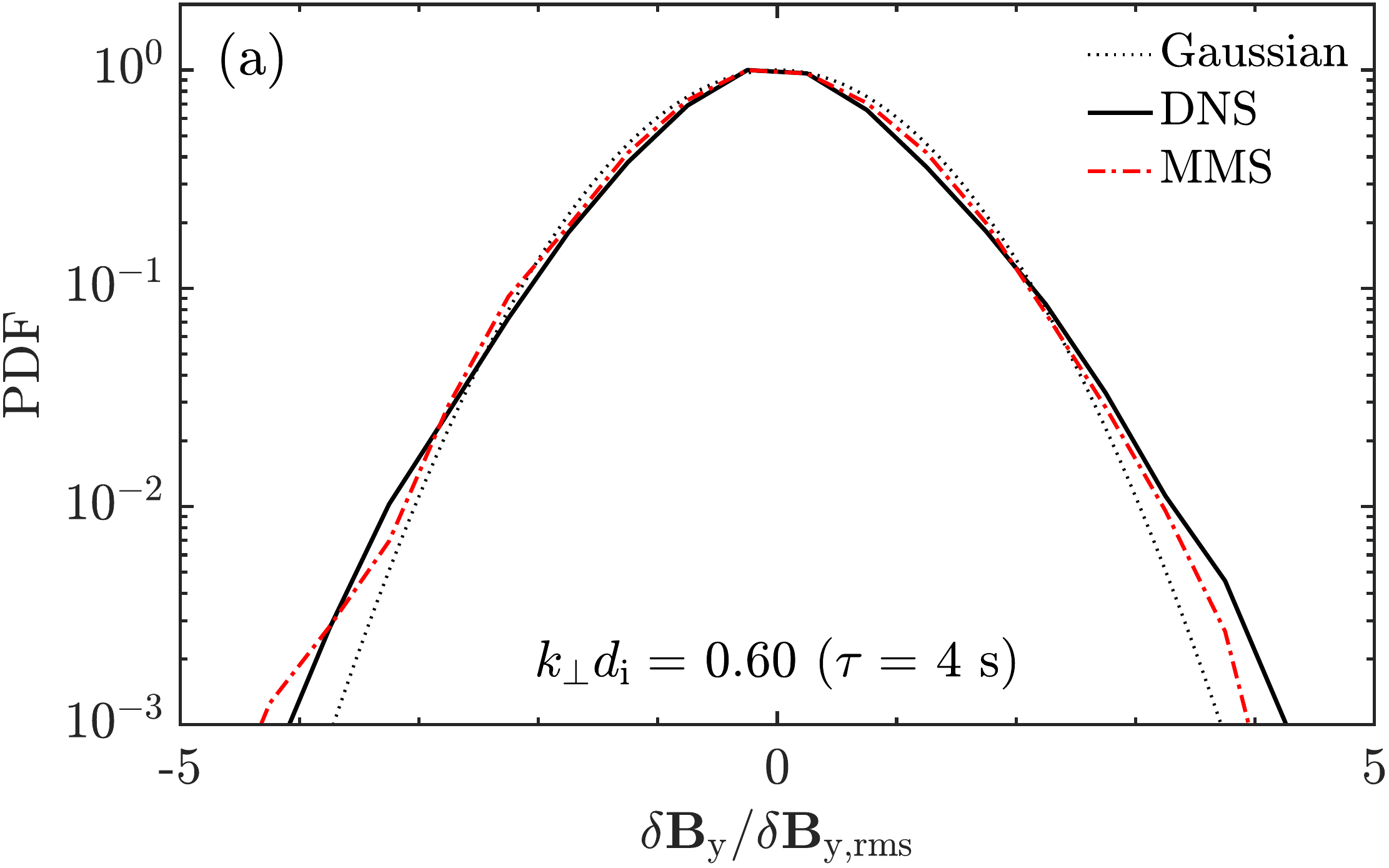}
\includegraphics[trim={0 1cm 0 0},clip,width=0.4\linewidth]{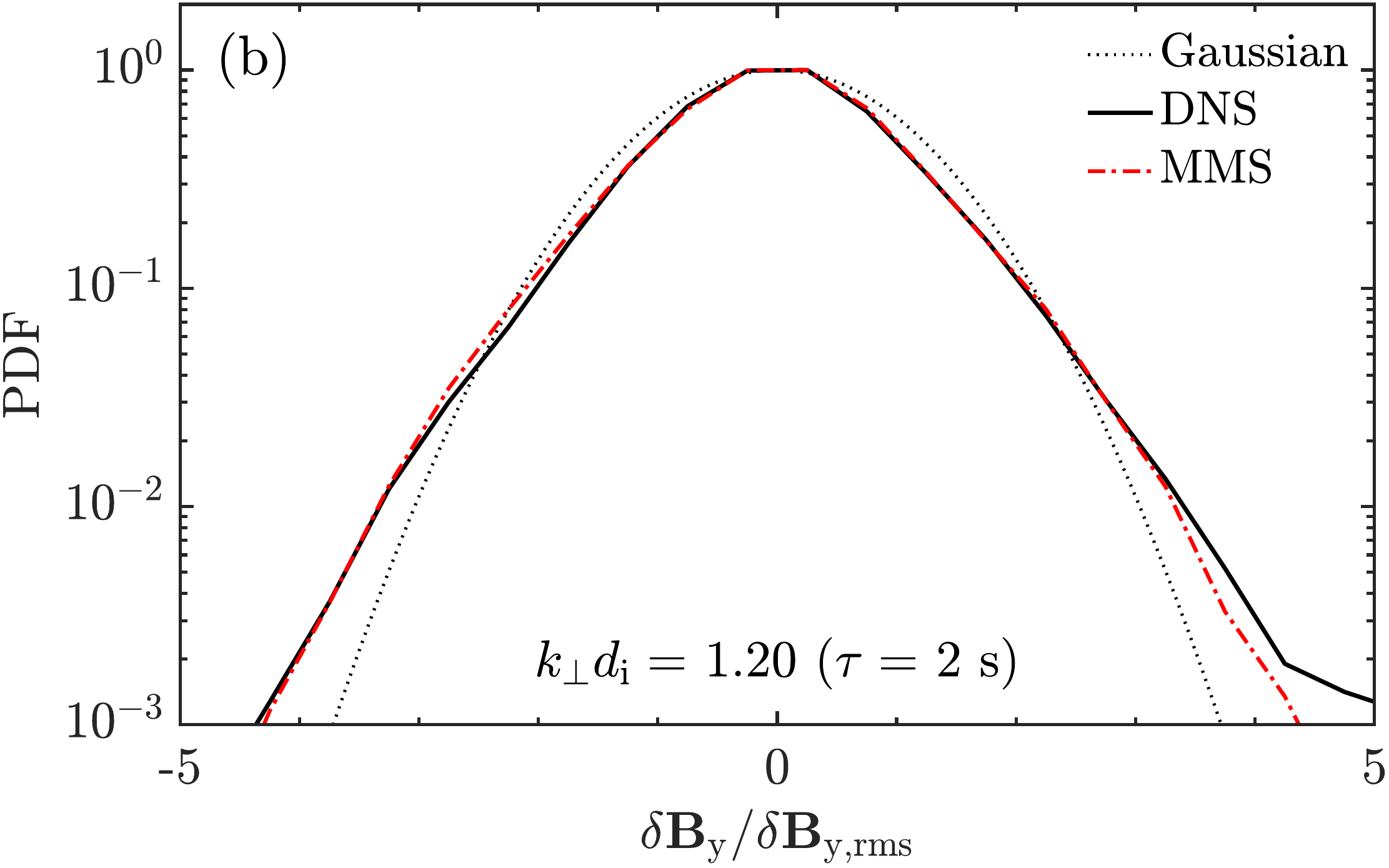}\\
\includegraphics[trim={0 1cm 0 0},clip,width=0.4\linewidth]{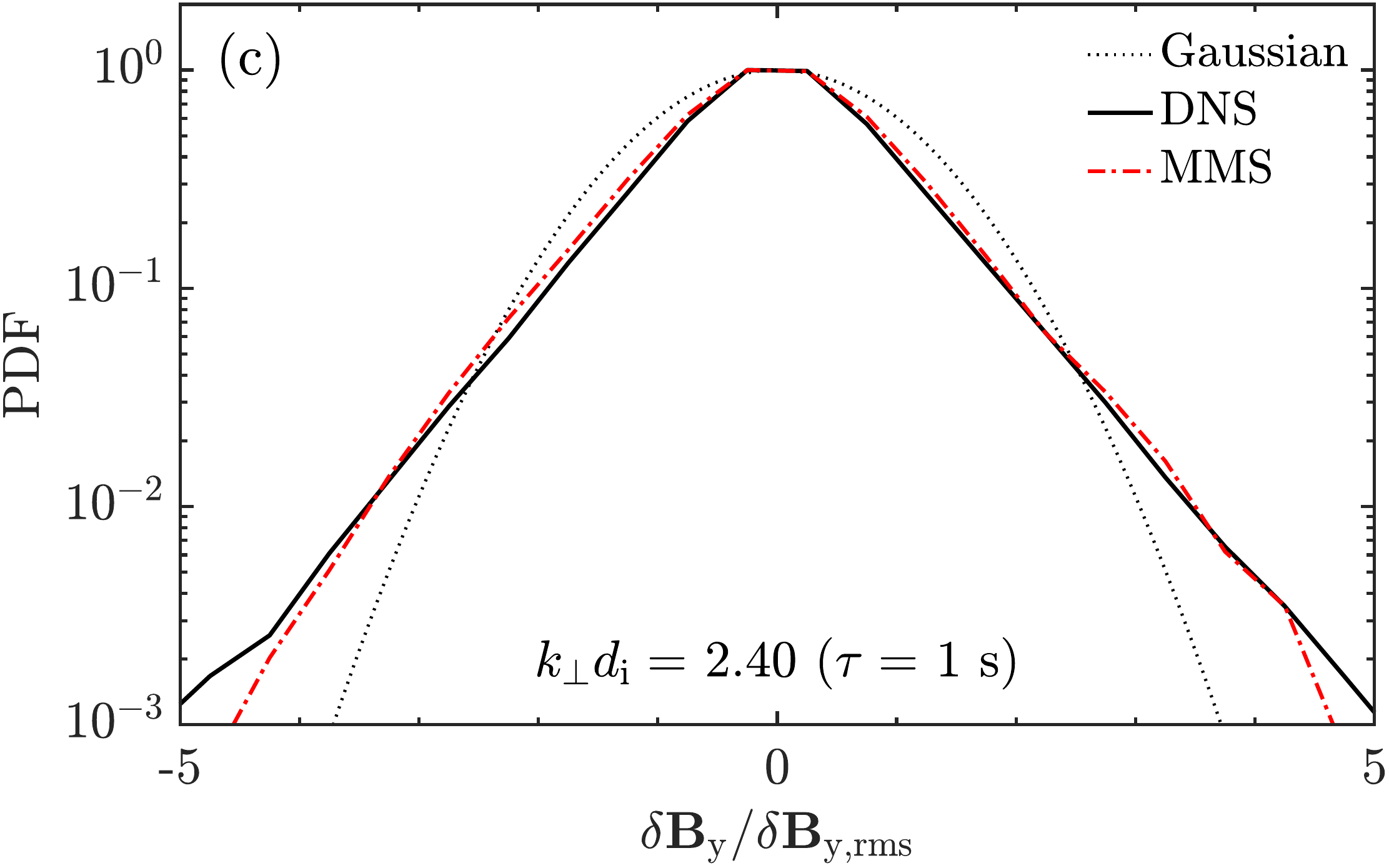}
\includegraphics[trim={0 1cm 0 0},clip,width=0.4\linewidth]{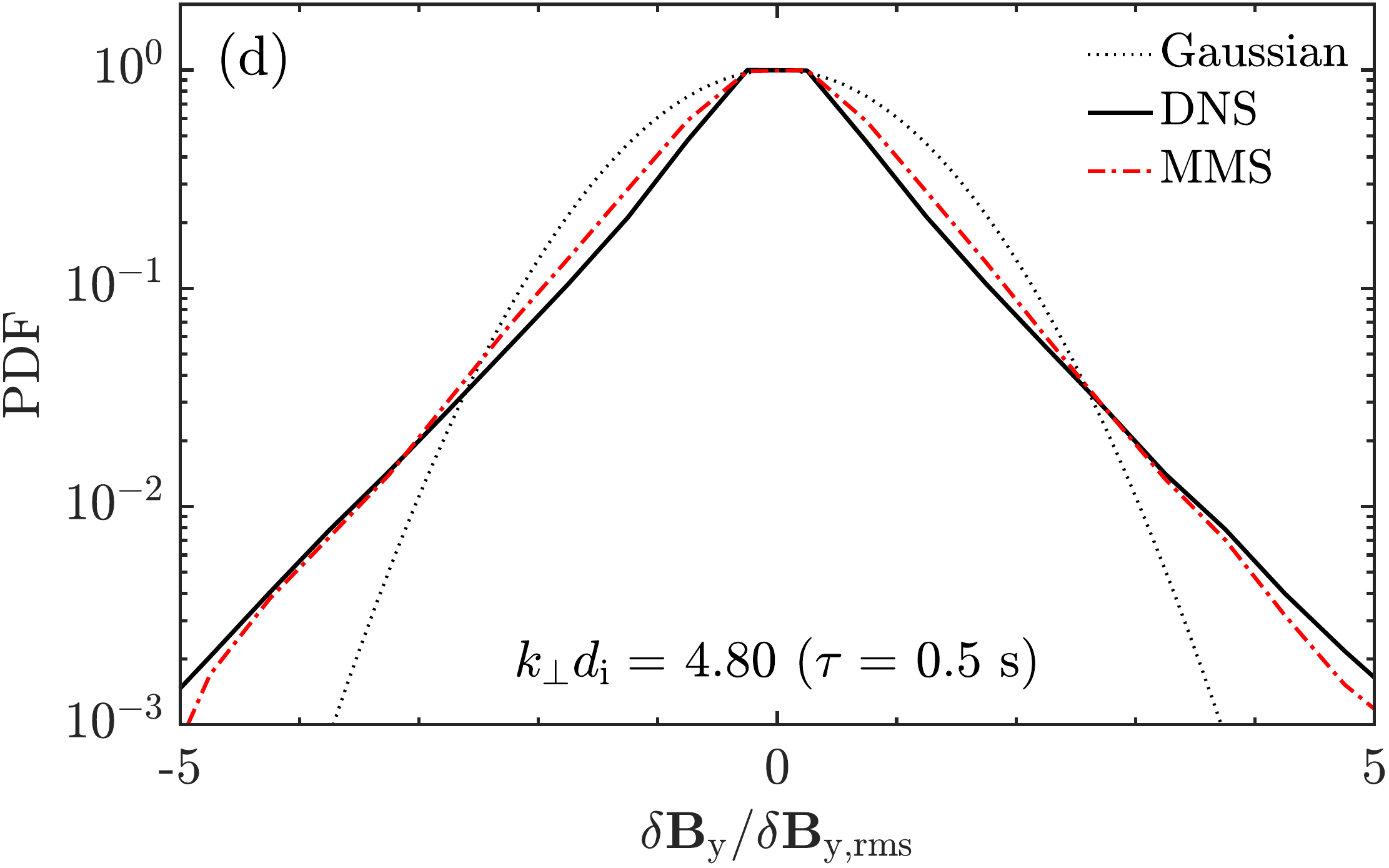}\\
\includegraphics[width=0.4\linewidth]{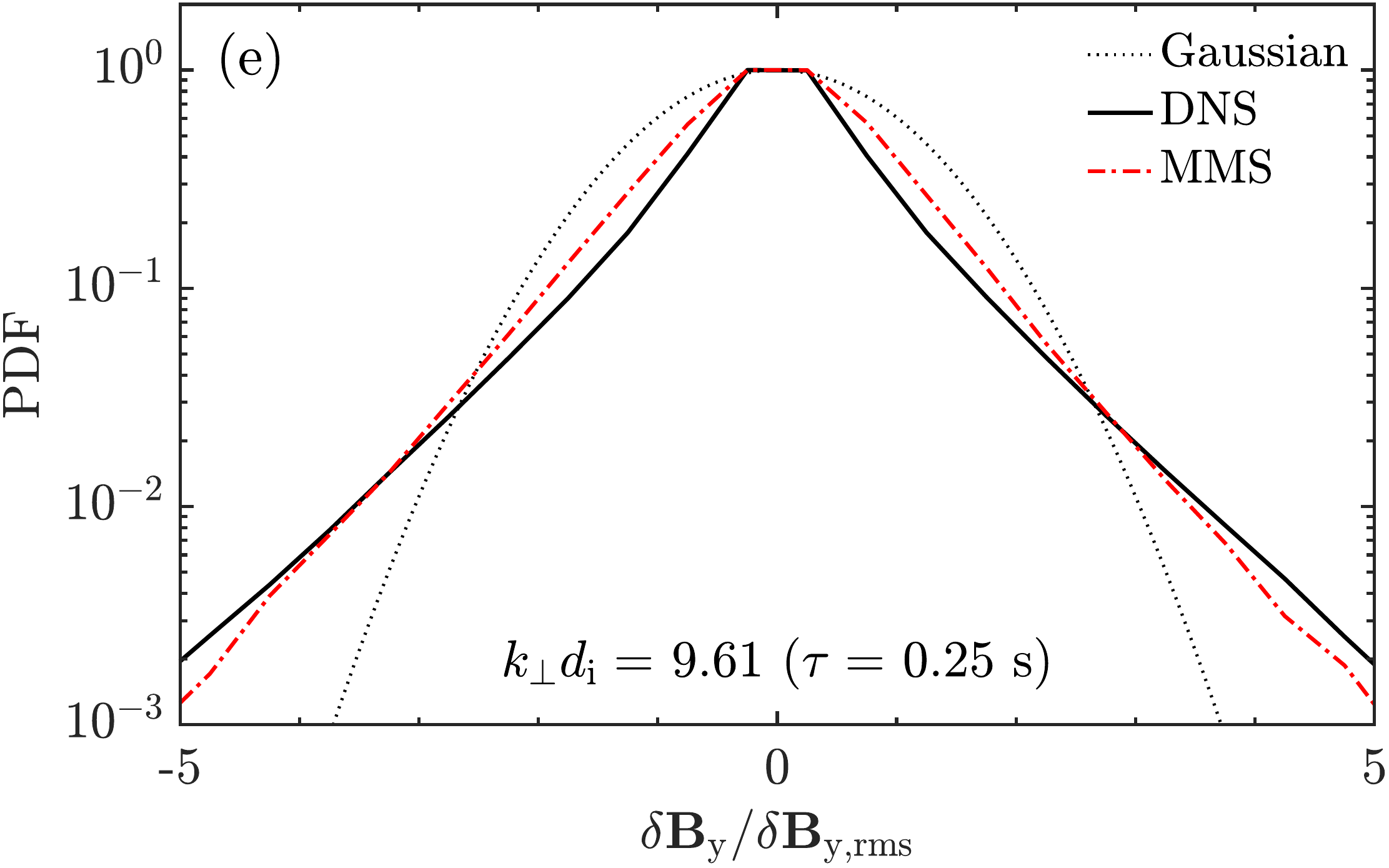}
\includegraphics[width=0.4\linewidth]{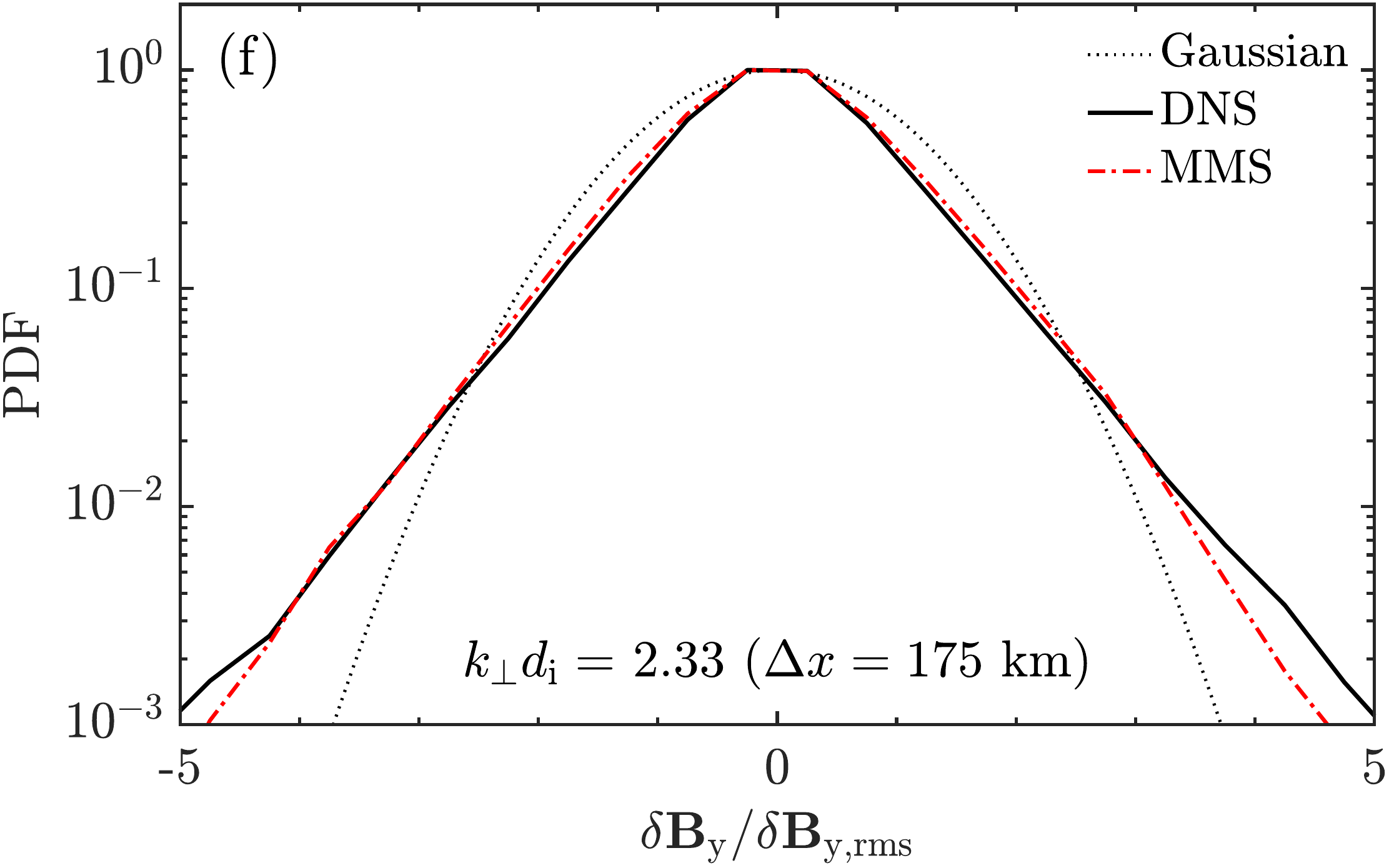}\\
\end{center}
\caption{(a-e) Normalized probability distribution functions of magnetic field increments at different wavenumbers, corresponding to different temporal lags.
  (f) The same as in the previous panels, for a spatial lag $\Delta x \approx 175$ km (corresponding to the MMS formation size).}
\label{fig:PDFs}
\end{figure*}
In the inertial range, the MMS and DNS density fluctuations exhibit a very different spectral behavior, due to the fact that the latter is initialized with no density fluctuations and this keeps the large-scale compressibility lower. The MMS spectrum exhibits a power-law behavior with a slope compatible with -1, consistent with previous spacecraft observation in the solar wind~\citep{Safrankova_al_2015}. At ion and sub-ion scales, a small difference in the level of fluctuations between MMS and the DNS is observed, of the same order of the one observed for the electron velocity spectra. Nevertheless, a good agreement is recovered for the slope below the ion scales, which is consistent with $-2.8$. This confirms that the level of compressibility in the inertial range does not impact significantly the nature of the fluctuations in the kinetic range~\citep{Cerri_al_2017a}.
The shape of the MMS spectrum of $n_{\rm{i}}$ (not shown) is the same as for $n_{\rm{e}}$ in the whole range of resolved scales. Its level in the kinetic range is slightly different, likely due to the MMS FPI density estimates, and in quantitative agreement with the DNS spectrum. 

The slopes of the different fields in the sub-ion range, i.e., $3 \lesssim k_\perp d_i \lesssim 40$, are not all independent. The one of the electric field, $\alpha_{\bm{E}} \sim - 0.8$, is related to the one of the density and of the parallel magnetic fluctuations (not shown), $\alpha_n \sim \alpha_{\bm{B}_{\parallel}} \sim - 2.8$, by the simple relation $\alpha_{\bm{E}} \sim \alpha_{n(\bm{B}_{\parallel})} -2$. This comes from the electron pressure gradient term and the Hall term in the generalized Ohm's law, the latter being dominated by its first-order contribution $(\nabla \times \bm{B}_\parallel) \times \bm{B}_0$ ~\citep{Franci_al_2015b}. The kinetic-scale slope of the electron bulk velocity spectrum is instead related to the one of the perpendicular magnetic fluctuations (not shown, but similar to the total magnetic field, representing its dominant contribution), $\alpha_{\bm{B}_{\perp}} \sim - 3.2$, according to $\alpha_{\bm{U}_e} \sim \alpha_{\bm{B}_{\perp}} + 2$. This is due to the fact that the out-of-plane current density is much larger than its in-plane component and the ion bulk velocity drops at ion scales, so $\bm{J} \sim \bm{J}_\parallel \sim \nabla \times \bm{B}_\perp \sim n_0 \bm{U}_e$.

\subsection{Intermittency}
\label{subsec:intermittency}

\begin{figure}[ht!]
\begin{center}
\includegraphics[width=\linewidth]{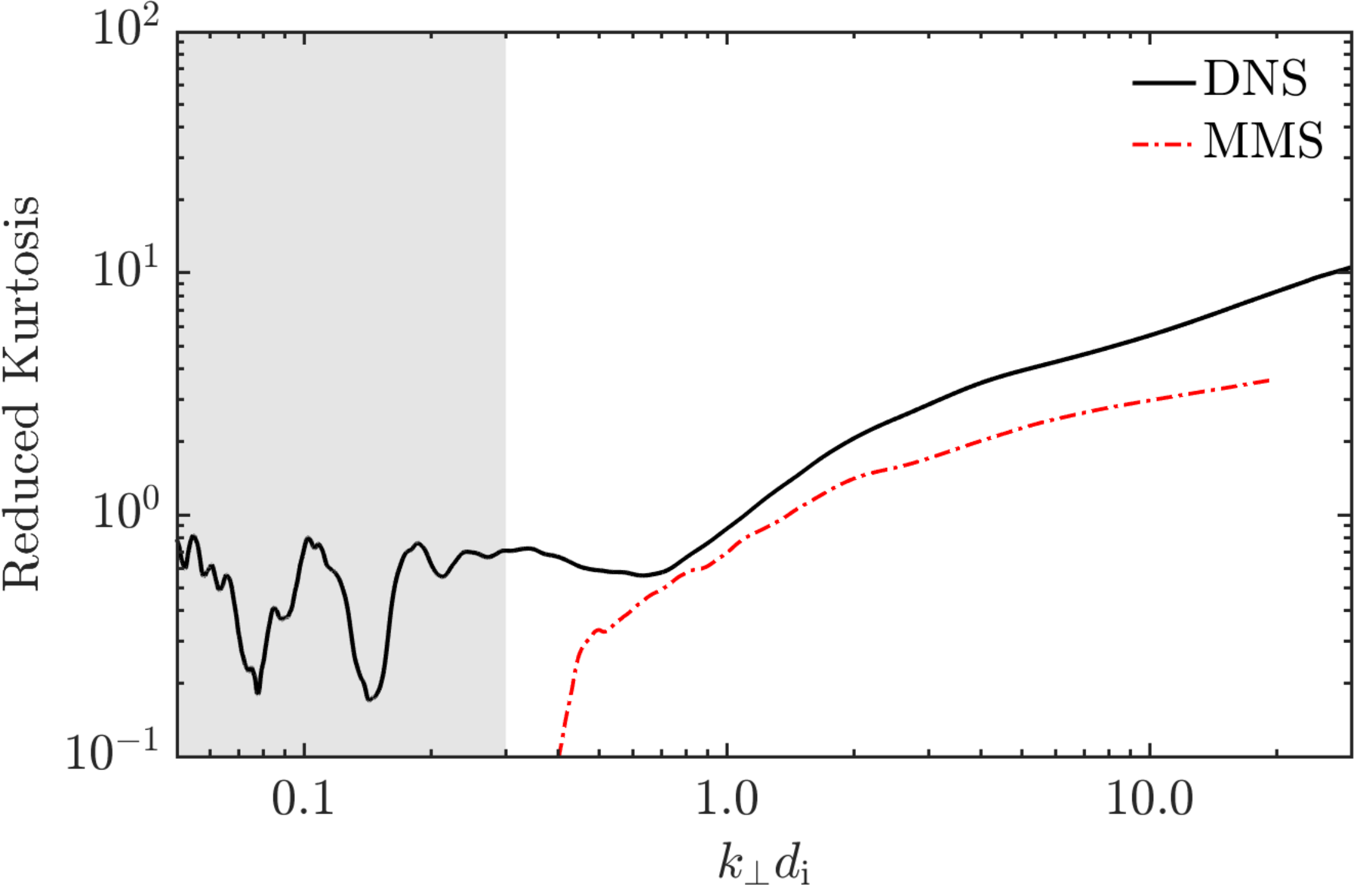}
\end{center}
\caption{Excess kurtosis of magnetic field increments. The grey shaded area marks the range of scales where enery is initially injected in the DNS.}
\label{fig:kurtosis}
\end{figure}

The intermittency properties are compared in Fig.~\ref{fig:PDFs}, showing the
PDFs of the increments of a
magnetic field component perpendicular to $\bm{B}_0$,
along the other perpendicular direction, normalized to its rms value, $\delta \mathbf{B}_{\mathrm{y}}(k_\perp) / \delta \mathbf{B}_{\mathrm{y,rms}}$. Six different values of $k_\perp d_i$ are explored, representing the inertial range, $k_\perp d_i = 0.6$ (panel (a)) and $1.2$ (b),
the ion-scale transition, $k_\perp d_i = 2.4$ (c) and $2.33$ (f), 
and the ion kinetic range, $k_\perp d_i = 4.8$ (d) and $9.6$ (e). 
For the DNS (black curves), these are simply obtained from
$k_\perp = 2\pi/\ell$, where $\ell$ is the spatial lag
in the simulation plane. For MMS (red curves), this only holds
for panel (f), where the increments are actually computed from the spatial
lag, while for (a)-(e) the temporal lag,
$\tau$, has been converted to the corresponding
wavenumber using the Taylor hypothesis. 

The PDFs of magnetic fluctuations present very
different shapes at different scales. At scales close to the injection scale, it is quite similar to a Gaussian distribution
(dotted black curve), deviating from it only in correspondence of the largest fluctuations. The PDFs become less and less Gaussian towards
smaller lags, developing extended tails at kinetic scales. 
For fluctuations
with an amplitude smaller than 4 times the rms value,
$\delta \mathbf{B}_{\mathrm{y}}(k_\perp) < 4 \, \delta \mathbf{B}_{\mathrm{y,rms}}$, the numerical and observational PDFs are almost overlapped at all scales, apart from small differences in the kinetic range.
In the tails, the PDFs from the DNS are usually higher than their observational counterpart (such differences are almost negligible if we consider that there the number of
counts is small and concentrated in very few points in the simulation
domain). Such intermittency properties are consistent with the excess kurtosis, shown in Fig.~\ref{fig:kurtosis}. Both the DNS (black) and the MMS (red) kurtosis are very small at large scales, down to $k_\perp \, d_i \sim 1$. They increase in the kinetic range, with a similar but still different trend, such that their values at $k_\perp \, d_i \sim 10$ differ by a factor of $\sim 2$. The DNS kurtosis is seen to be very sensitive to the tails of the PDFs, where the results suffer from resolution effects. Moreover, at earlier times, it assumes larger values of about 2 orders of magnitude, peaking at more than 100 in correspondence with the onset of magnetic reconnection, when strong current sheets are just about to start being disrupted.

\subsection{Cascade rate}
\label{subsec:yaglom}

\begin{figure}[ht!]
\begin{center}
\includegraphics[width=\linewidth]{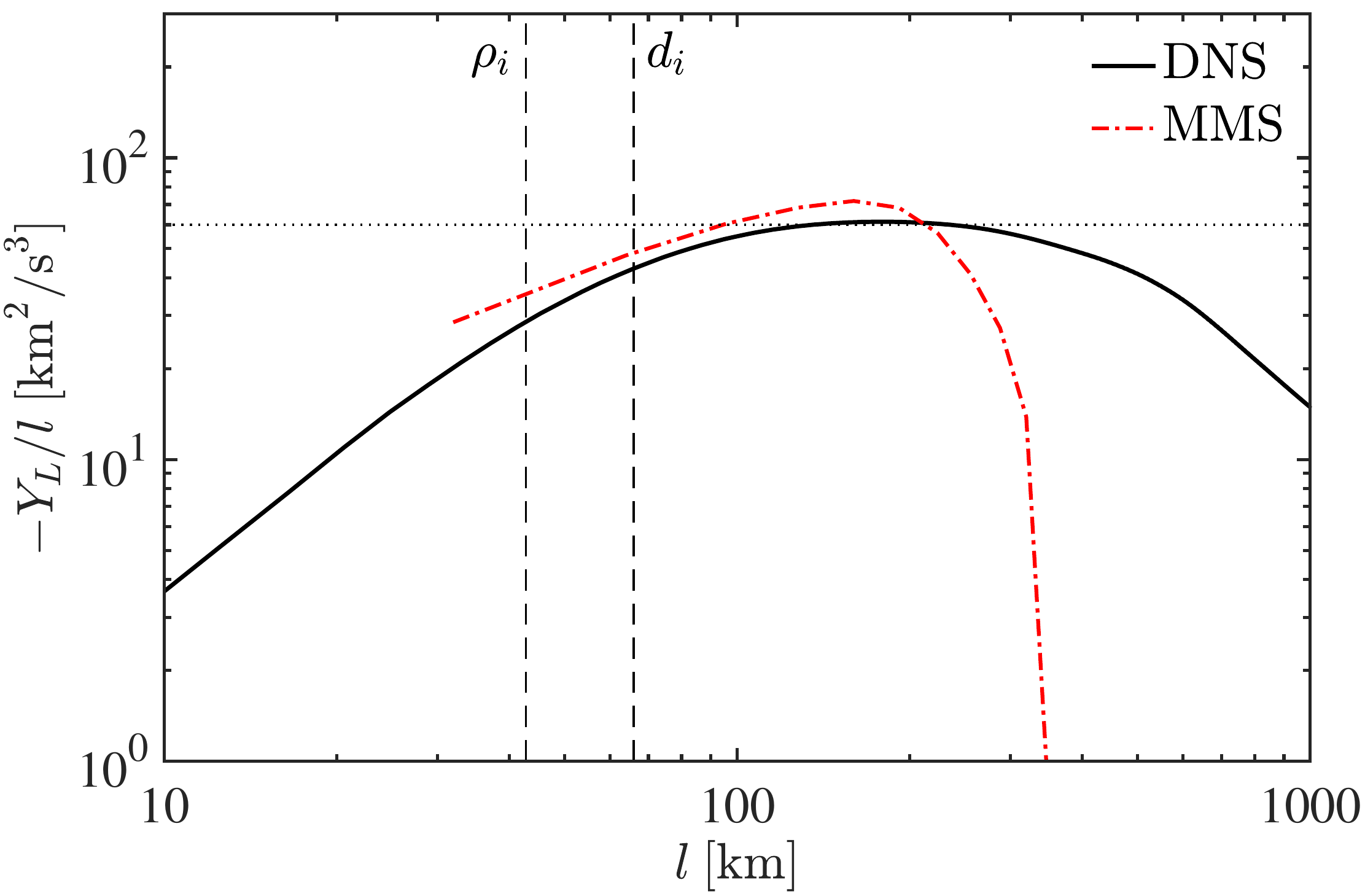}\\
\end{center}
\caption{MHD structure function from generalized third-order law,
  $-\bm{Y}$, divided by the spatial lag, $l$, versus $l$ in km.}
\label{fig:Yaglom}
\end{figure}

The cascade rate properties can be investigated by using the statistical von K\'arm\'an-Howarth/Politano-Pouquet law \citep{Karman_Howarth_1938,Politano_Pouquet_1998} for the mixed 3rd-order structure function
\begin{equation}
 \boldsymbol{Y}=\left\langle
\delta\boldsymbol{U}\left|\delta\boldsymbol{U}\right|^{2}+\delta\boldsymbol{U}\left|\delta\boldsymbol{b}\right|^{2}-2\delta\boldsymbol{b}\left(\delta\boldsymbol{U}\cdot\delta\boldsymbol{b}\right)\right\rangle,
 \label{Y}
\end{equation}
where 
the increments of the bulk velocity, $\delta \boldsymbol{U}$, and of the magnetic field
in Alfv\'enic units, $\delta \boldsymbol{b}$, are
computed between two points
$x$ and  $x + l$ and $\langle \; \rangle$ denotes the average over $x$. This law predicts for the MHD inertial range
a linear scaling of $Y_L$ with the separation scale $l$, where $Y_L$ is longitudinal component.
Figure~\ref{fig:Yaglom} compares $-Y_L/l$ as a function of $l$ from MMS and the DNS.
Both curves exhibit a similar profile with a plateau, but they do not
show a clear linear scaling of $Y_L$ (or a constant value of $-Y_L/l$), a signature of
the inertial range. This is likely due to the fact that the energy injection
takes place at scales not far enough from the ion ones and in the sub-ion range
the Hall physics must be included \citep{Papini_al_2019}.
Indeed, recent simulations and observations indicate that
the turbulent energy cascade partly continues via the Hall
term \citep{Hellinger_al_2018,Bandyopadhyay_al_2019}.
Figure~\ref{fig:Yaglom}, however, shows that
MMS and the DNS exhibit very similar values (for the
separation length in the range
between about 50 to 200 km). Assuming isotropy, we get (at the plateau)
a cascade rate of $\sim 5 \times 10^7 \mathrm{J/kg/s}$, i.e., a value about
10 times larger than the one observed in the Earth's magnetosheath
\citep{Bandyopadhyay_al_2018,Hadid_al_2018}, consistent with the turbulence being driven by the KH instability and not just pre-existing background turbulence.

\section{Conclusions and discussion}
\label{sec:discussion}

We have reproduced MMS observations of KH-generated turbulence in the Earth's magnetopause with a high-resolution 2D hybrid DNS of Alfv\'enic plasma turbulence. We have used the observed values of four fundamental physical parameters determining the plasma conditions (ion and electron beta, turbulence strength, and injection scale). No other free parameters have had to be used to optimize the agreement.

The MMS and DNS spectra of the electromagnetic fluctuations exhibit an unprecedented quantitative agreement. The ion beta controls the scale and shape of the ion-scale spectral break in the magnetic field~\citep{Franci_al_2016b}, while the electron beta affects the level of the electron-pressure-gradient term in the generalized Ohm's law~\citep{Franci_al_2015b,Matteini_al_2017}, determining the level of electric field fluctuations in the inertial range. 
Although the level of ion and electron bulk velocity and density fluctuations is slightly different, the spectral indices are very similar. The only major difference is the level of compressibility at MHD scales, lower in the DNS due to the initial conditions. 

The agreement between MMS observations and the DNS extends to intermittency. In both cases the PDFs of magnetic fluctuations in the inertial range are very close to a Gaussian distribution and, consequently, the excess kurtosis is close to zero, rapidly increasing monotonically towards smaller scales. At both extremes of the range of scales here investigated, the intermittency properties differ from what typically observed in the solar wind, where the kurtosis is  larger at MHD scales and exhibits a plateau or a decrease at sub-ion scales~\citep[e.g.,][]{Kiyani_al_2009,Kiyani_al_2013,Wu_al_2013,Chen_al_2014a,Chen_2016,Chhiber_al_2018}.
The former discrepancy is related to the fact that the injection scale here is just about one order of magnitude larger than the ion scales, so that MHD-scale current sheets are unlikely to form and only vortices (and, possibly, wave-like fluctuations) are present in the inertial range.  The origin of the different behavior at sub-ion scales is not trivial. It could be related to a larger presence of proton and electron-scale current sheets in the magnetosheath with respect to the solar wind, as shown in \cite{Chhiber_al_2018}, possibly as the consequence of the continuous formation of KH vortices interacting with each other. This topic will be further discussed in a following paper.


The behavior of the mixed 3rd-order structure function is consistent with the picture above: both the MMS and the DNS inertial ranges are quite small and a change in the nature of the fluctuations is observed at ion scales. This is where the MHD approximation of the 3rd-order exact law stops to hold and Hall and kinetic effects need to be taken into account to have a comprehensive estimate of the cascade rate. The KH instability-driven turbulence 
is observed to be much stronger than the ambient magnetosheath turbulence, since its cascade rate is an order of magnitude larger.

The overall quantitative agreement between MMS observations and our DNS confirms that the hybrid approach is optimal for investigating the properties of the turbulent cascade simultaneously from MHD down to sub-ion scales, provided that a high accuracy is employed. All the main physical processes seem to be included and correctly reproduced in the model and the ones missing are likely not fundamental in the plasma dynamics at these scales and in these plasma conditions. As a consequence, as far as the spectral and intermittency properties are concerned, they represent a preferable choice in those situations where the cascade needs to be followed over a wide range of scales simultaneously and/or with a very high accuracy at kinetic scales.

This work further confirms the quasi-2D nature of plasma turbulence in the presence of a strong guide field, with the cascade mainly developing in the perpendicular direction, as suggested by 3D DNS employing different methods ~\citep[e.g.][and references therein]{Cerri_al_2019} and by a direct quantitative comparison between 2D and 3D CAMELIA DNS~\citep{Franci_al_2018a,Franci_al_2018b}. 

The Taylor hypothesis is shown to hold even in the magnetopause and down to sub-ion scales~\citep{Chen_al_2017,Stawarz_al_2019,Chhiber_al_2019}, implying that the fluctuations at sub-ion scales must be predominantly low-frequency, consistent with quasi-static structures or with a Kinetic Alfv\'en wave-like nature rather than whistler. 

Interestingly, in both the KH event and in the DNS a strong interplay between turbulence and magnetic reconnection is observed. In the magnetopause, reconnection occurs in intense current sheets forming at the trailing edges of KH-related surface waves and turbulence develops in the reconnection exhausts~\citep{Eriksson_al_2016a}. In the DNS, reconnection is driven by the interaction of large-scale vortices and it is observed to act as a driver for the development of the kinetic-scale cascade~\citep{Franci_al_2017,Papini_al_2019}.
Since the properties of turbulence and reconnection, as well as their interplay, are likely controlled by the plasma conditions (including the injection scale, rather than by the nature of the injection mechanism), the role of reconnection observed in both the KH event and in the DNS could provide a possible explanation for the agreement. Indeed, there is evidence that the correlation length of the turbulence can change how reconnection proceeds (i.e. electron-only reconnection)~\citep{Phan_al_2018,Stawarz_al_2019,SharmaPyakurel_al_2019}. 

Summarising, the  physical implications of the direct comparison between MMS observation and our DNS are:
\begin{enumerate}[label=(\roman*)]
\item KH instability-driven turbulence in the magnetopause has similar spectral and intermittency properties compatible with Alfv\'enic turbulence from MHD scales down to sub-ion scales;
\item the plasma dynamics is controlled by a few fundamental plasma parameters, and the energy injection scale is much more important than the nature of the driving mechanism itself, hinting at a certain degree of universality;
\item the main properties of the fluctuations (e.g., compressibility) at ion and sub-ion scales are independent on the inertial range, possibly suggesting -together with point (ii)- a certain degree of universality of the kinetic turbulent cascade;
\item electron kinetic processes, intrinsically absent in the DNS, do not seem to play a significant role for the properties here compared at scales larger than the electron characteristic scales, in the particular investigated regime (intermediate ion beta and low electron beta);
\item fluctuations at ion and sub-ion scales are likely low-frequency, consistent with a kinetic Alfv\'en wave-like nature or with quasi-static structures;
\item our DNS results represent a good model for the particular observed event, compatible with a quasi-2D nature of the turbulent cascade; 
\item the inertial-range intermittency is smaller than in the pristine solar wind, consistent with a smaller correlation length of the turbulence due to energy injection at scales closer to the ion scales; 
\item the kurtosis does not saturate at sub-ion scales in the magnetopause, possibly due to a different contribution from coherent structures and/or waves and phase-randomizing structures with respect to the solar wind; 
\item the larger cascade rate than the one measured in the ambient magnetosheath suggests that the turbulence we are observing is indeed driven by the KH event rather than pre-existing turbulence that the HK instability occurs on top of.
\end{enumerate}

A key difference of our DNS with respect to observations and other simulations investigating the same event (Nakamura et al, submitted) is the initial velocity. The KH instability becomes unstable when the velocity shear between the solar wind and the magnetosphere exceeds the Alfv\'en speed. Hence, the vortex-induced reconnection (VIR) process is controlled by the super-Alfv\'enic vortex flow \citep[e.g.][]{Nakamura_al_2011,Nakamura_al_2013}. As a consequence, the first injection to the VIR turbulence is caused by the strong background plasma flows, that are not included in the present DNS. 
Although the focus of this study is the stage after the first injection and the consequent development of the turbulent cascade, including realistic injection mechanisms, e.g. the KH instability, might provide complementary information. 

Compressible fluctuations could also be included in the initialization, providing for a better modeling in the interial range. Despite these likely play a role in the large-scale dynamics, there is evidence that the kinetic-scale compressibility is quite independent on its MHD-scale counterpart~\citep[e.g.,][]{Cerri_al_2017a}.

Full kinetic simulations will allow us to extend our modeling of the observed turbulent properties down to electron characteristic scales.


%

\acknowledgments
The authors wish to acknowledge valuable discussions with Christopher Chen, William H. Matthaeus, Daniel Verscharen, Riddhi Bandyopadhyay, Leonardo Bianchini, and Christopher Callow. L.F. and D.B. are supported by  UK Science and Technology Facilities Council (STFC) grant ST/P000622/1. J.E.S. is supported by STFC grant ST/S000364/1. P.H. acknowledges grant 18-08861S of the Czech Science Foundation. 
This research utilised Queen Mary's Apocrita HPC facility, supported by
QMUL Research-IT. http://doi.org/10.5281/zenodo.438045. We acknowledge
PRACE for awarding us access to resource Cartesius based in the
Netherlands at SURFsara through the DECI-13 (Distributed European
Computing Initiative) call (project HybTurb3D), and the CINECA award
under the ISCRA initiative, for the availability of high performance
computing resources and support (grants HP10CQI4XP and
HP10B2DRR4). The LPP involvement for the SCM instrument is supported by CNES and CNRS.
The observational data are publicly available through the MMS Science Data Center (https://lasp.colorado.edu/mms/sdc/public/).


\begin{thebibliography}{}
\expandafter\ifx\csname natexlab\endcsname\relax\def\natexlab#1{#1}\fi
\providecommand{\url}[1]{\href{#1}{#1}}

\bibitem[{{Alexandrova} {et~al.}(2009){Alexandrova}, {Saur}, {Lacombe},
  {Mangeney}, {Mitchell}, {Schwartz}, \& {Robert}}]{Alexandrova_al_2009}
{Alexandrova}, O., {Saur}, J., {Lacombe}, C., {et~al.} 2009, Phys.~Rev.~Lett.,
  103, arXiv:0906.3236

\bibitem[{Arzamasskiy {et~al.}(2019)Arzamasskiy, Kunz, Chandran, \&
  Quataert}]{Arzamasskiy_al_2019}
Arzamasskiy, L., Kunz, M.~W., Chandran, B. D.~G., \& Quataert, E. 2019, The
  Astrophysical Journal, 879, 53.
\newblock \url{https://doi.org/10.3847%2F1538-4357%2Fab20cc}

\bibitem[{{Bale} {et~al.}(2005){Bale}, {Kellogg}, {Mozer}, {Horbury}, \&
  {Reme}}]{Bale_al_2005}
{Bale}, S.~D., {Kellogg}, P.~J., {Mozer}, F.~S., {Horbury}, T.~S., \& {Reme},
  H. 2005, Phys.~Rev.~Lett., 94, 215002

\bibitem[{{Bandyopadhyay} {et~al.}(2018){Bandyopadhyay}, {Chasapis}, {Chhiber},
  {Parashar}, {Matthaeus}, {Shay}, {Maruca}, {Burch}, {Moore}, {Pollock},
  {Giles}, {Paterson}, {Dorelli}, {Gershman}, {Torbert}, {Russell}, \&
  {Strangeway}}]{Bandyopadhyay_al_2018}
{Bandyopadhyay}, R., {Chasapis}, A., {Chhiber}, R., {et~al.} 2018,
  Astrophys.~J., 866, 106

\bibitem[{{Bandyopadhyay} {et~al.}(2019){Bandyopadhyay}, {Sorriso-Valvo},
  {Chasapis}, {Hellinger}, {Matthaeus}, {Verdini}, {Landi}, {Franci},
  {Matteini}, {Giles}, {Pollock}, {Russell}, {Strangeway}, {Torbert}, {Moore},
  \& {Burch}}]{Bandyopadhyay_al_2019}
{Bandyopadhyay}, R., {Sorriso-Valvo}, L., {Chasapis}, A.~r., {et~al.} 2019,
  arXiv e-prints, arXiv:1907.06802

\bibitem[{{Boldyrev} {et~al.}(2013){Boldyrev}, {Horaites}, {Xia}, \&
  {Perez}}]{Boldyrev_al_2013}
{Boldyrev}, S., {Horaites}, K., {Xia}, Q., \& {Perez}, J.~C. 2013,
  Astrophys.~J., 777, 41

\bibitem[{{Boldyrev} {et~al.}(2011){Boldyrev}, {Perez}, {Borovsky}, \&
  {Podesta}}]{Boldyrev_al_2011}
{Boldyrev}, S., {Perez}, J.~C., {Borovsky}, J.~E., \& {Podesta}, J.~J. 2011,
  Astrophys. J. Let., 741, arXiv:1106.0700

\bibitem[{{Breuillard} {et~al.}(2018){Breuillard}, {Matteini}, {Argall},
  {Sahraoui}, {Andriopoulou}, {Le Contel}, {Retin{\`o}}, {Mirioni}, {Huang},
  {Gershman}, {Ergun}, {Wilder}, {Goodrich}, {Ahmadi}, {Yordanova}, {Vaivads},
  {Turner}, {Khotyaintsev}, {Graham}, {Lindqvist}, {Chasapis}, {Burch},
  {Torbert}, {Russell}, {Magnes}, {Strangeway}, {Plaschke}, {Moore}, {Giles},
  {Paterson}, {Pollock}, {Lavraud}, {Fuselier}, \&
  {Cohen}}]{Breuillard_al_2018}
{Breuillard}, H., {Matteini}, L., {Argall}, M.~R., {et~al.} 2018, \apj, 859,
  127

\bibitem[{Bruno(2019)}]{Bruno_2019}
Bruno, R. 2019, Earth and Space Science, 6, 656.
\newblock
  \url{https://agupubs.onlinelibrary.wiley.com/doi/abs/10.1029/2018EA000535}

\bibitem[{{Bruno} \& {Carbone}(2013)}]{Bruno_Carbone_2013}
{Bruno}, R., \& {Carbone}, V. 2013, Living Rev. Solar Phys., 10, 2

\bibitem[{{Cerri} {et~al.}(2017){Cerri}, {Franci}, {Califano}, {Landi}, \&
  {Hellinger}}]{Cerri_al_2017a}
{Cerri}, S.~S., {Franci}, L., {Califano}, F., {Landi}, S., \& {Hellinger}, P.
  2017, J.~Plasma Phys., 83, 705830202

\bibitem[{Cerri {et~al.}(2019)Cerri, Grošelj, \& Franci}]{Cerri_al_2019}
Cerri, S.~S., Grošelj, D., \& Franci, L. 2019, Frontiers in Astronomy and
  Space Sciences, 6, 64.
\newblock \url{https://www.frontiersin.org/article/10.3389/fspas.2019.00064}

\bibitem[{{Chandran} {et~al.}(2009){Chandran}, {Quataert}, {Howes}, {Xia}, \&
  {Pongkitiwanichakul}}]{Chandran_al_2009}
{Chandran}, B.~D.~G., {Quataert}, E., {Howes}, G.~G., {Xia}, Q., \&
  {Pongkitiwanichakul}, P. 2009, Astrophys. J., 707, 1668

\bibitem[{{Chen}(2016)}]{Chen_2016}
{Chen}, C.~H.~K. 2016, J.~Plasma Phys., 82, 535820602

\bibitem[{{Chen} \& {Boldyrev}(2017)}]{Chen_al_2017}
{Chen}, C.~H.~K., \& {Boldyrev}, S. 2017, \apj, 842, 122

\bibitem[{{Chen} {et~al.}(2013){Chen}, {Boldyrev}, {Xia}, \&
  {Perez}}]{Chen_al_2013}
{Chen}, C.~H.~K., {Boldyrev}, S., {Xia}, Q., \& {Perez}, J.~C. 2013,
  Phys.~Rev.~Lett., 110, arXiv:1305.2950

\bibitem[{{Chen} {et~al.}(2012){Chen}, {Salem}, {Bonnell}, {Mozer}, \&
  {Bale}}]{Chen_al_2012}
{Chen}, C.~H.~K., {Salem}, C.~S., {Bonnell}, J.~W., {Mozer}, F.~S., \& {Bale},
  S.~D. 2012, Phys.~Rev.~Lett., 109, arXiv:1205.5063

\bibitem[{{Chen} {et~al.}(2014){Chen}, {Sorriso-Valvo}, {{\v
  S}afr{\'a}nkov{\'a}}, \& {N{\v e}me{\v c}ek}}]{Chen_al_2014a}
{Chen}, C.~H.~K., {Sorriso-Valvo}, L., {{\v S}afr{\'a}nkov{\'a}}, J., \& {N{\v
  e}me{\v c}ek}, Z. 2014, Astrophys. J. Let., 789, arXiv:1405.7189

\bibitem[{{Chhiber} {et~al.}(2019){Chhiber}, {Usmanov}, {Matthaeus},
  {Parashar}, \& {Goldstein}}]{Chhiber_al_2019}
{Chhiber}, R., {Usmanov}, A.~V., {Matthaeus}, W.~H., {Parashar}, T.~N., \&
  {Goldstein}, M.~L. 2019, Astrophys. J. Suppl. Ser., 242, 12

\bibitem[{Chhiber {et~al.}(2018)Chhiber, Chasapis, Bandyopadhyay, Parashar,
  Matthaeus, Maruca, Moore, Burch, Torbert, Russell, Le~Contel, Argall,
  Fischer, Mirioni, Strangeway, Pollock, Giles, \& Gershman}]{Chhiber_al_2018}
Chhiber, R., Chasapis, A., Bandyopadhyay, R., {et~al.} 2018, Journal of
  Geophysical Research: Space Physics, 123, 9941.
\newblock
  \url{https://agupubs.onlinelibrary.wiley.com/doi/abs/10.1029/2018JA025768}

\bibitem[{{de K\'arm\'an} \& {Howarth}(1938)}]{Karman_Howarth_1938}
{de K\'arm\'an}, T., \& {Howarth}, L. 1938, Proc. Royal Soc. London Series A,
  164, 192

\bibitem[{{Ergun} {et~al.}(2016){Ergun}, {Tucker}, {Westfall}, {Goodrich},
  {Malaspina}, {Summers}, {Wallace}, {Karlsson}, {Mack}, {Brennan}, {Pyke},
  {Withnell}, {Torbert}, {Macri}, {Rau}, {Dors}, {Needell}, {Lindqvist},
  {Olsson}, \& {Cully}}]{Ergun_al_2016}
{Ergun}, R.~E., {Tucker}, S., {Westfall}, J., {et~al.} 2016, Space Sci. Rev.,
  199, 167

\bibitem[{Eriksson {et~al.}(2016)Eriksson, Lavraud, Wilder, Stawarz, Giles,
  Burch, Baumjohann, Ergun, Lindqvist, Magnes, Pollock, Russell, Saito,
  Strangeway, Torbert, Gershman, Khotyaintsev, Dorelli, Schwartz, Avanov,
  Grimes, Vernisse, Sturner, Phan, Marklund, Moore, Paterson, \&
  Goodrich}]{Eriksson_al_2016a}
Eriksson, S., Lavraud, B., Wilder, F.~D., {et~al.} 2016, Geophys. Res. Lett.,
  doi:10.1002/2016GL068783

\bibitem[{{Eriksson} {et~al.}(2016){Eriksson}, {Wilder}, {Ergun}, {Schwartz},
  {Cassak}, {Burch}, {Chen}, {Torbert}, {Phan}, {Lavraud}, {Goodrich},
  {Holmes}, {Stawarz}, {Sturner}, {Malaspina}, {Usanova}, {Trattner},
  {Strangeway}, {Russell}, {Pollock}, {Giles}, {Hesse}, {Lindqvist}, {Drake},
  {Shay}, {Nakamura}, \& {Marklund}}]{Eriksson_al_2016b}
{Eriksson}, S., {Wilder}, F.~D., {Ergun}, R.~E., {et~al.} 2016,
  Phys.~Rev.~Lett., 117, 015001

\bibitem[{{Franci} {et~al.}(2018{\natexlab{a}}){Franci}, {Hellinger},
  {Guarrasi}, {Chen}, {Papini}, {Verdini}, {Matteini}, \&
  {Landi}}]{Franci_al_2018b}
{Franci}, L., {Hellinger}, P., {Guarrasi}, M., {et~al.} 2018{\natexlab{a}}, in
  Journal of Physics Conference Series, Vol. 1031, Journal of Physics
  Conference Series, 012002

\bibitem[{{Franci} {et~al.}(2015{\natexlab{a}}){Franci}, {Landi}, {Matteini},
  {Verdini}, \& {Hellinger}}]{Franci_al_2015b}
{Franci}, L., {Landi}, S., {Matteini}, L., {Verdini}, A., \& {Hellinger}, P.
  2015{\natexlab{a}}, Astrophys.~J., 812, 21

\bibitem[{{Franci} {et~al.}(2016){Franci}, {Landi}, {Matteini}, {Verdini}, \&
  {Hellinger}}]{Franci_al_2016b}
---. 2016, Astrophys.~J., 833, 91

\bibitem[{{Franci} {et~al.}(2018{\natexlab{b}}){Franci}, {Landi}, {Verdini},
  {Matteini}, \& {Hellinger}}]{Franci_al_2018a}
{Franci}, L., {Landi}, S., {Verdini}, A., {Matteini}, L., \& {Hellinger}, P.
  2018{\natexlab{b}}, Astrophys.~J., 853, 26

\bibitem[{{Franci} {et~al.}(2015{\natexlab{b}}){Franci}, {Verdini}, {Matteini},
  {Landi}, \& {Hellinger}}]{Franci_al_2015a}
{Franci}, L., {Verdini}, A., {Matteini}, L., {Landi}, S., \& {Hellinger}, P.
  2015{\natexlab{b}}, Astrophys. J. Let., 804, arXiv:1503.05457

\bibitem[{{Franci} {et~al.}(2017){Franci}, {Cerri}, {Califano}, {Landi},
  {Papini}, {Verdini}, {Matteini}, {Jenko}, \& {Hellinger}}]{Franci_al_2017}
{Franci}, L., {Cerri}, S.~S., {Califano}, F., {et~al.} 2017,
  Astrophys.~J.~Lett., 850, L16

\bibitem[{{Gary} {et~al.}(2012){Gary}, {Chang}, \& {Wang}}]{Gary_al_2012}
{Gary}, S.~P., {Chang}, O., \& {Wang}, J. 2012, Astrophys.~J., 755, 142

\bibitem[{{Goldreich} \& {Sridhar}(1995)}]{Goldreich_Sridhar_1995}
{Goldreich}, P., \& {Sridhar}, S. 1995, Astrophys. J., 438, 763

\bibitem[{Gro\ifmmode~\check{s}\else \v{s}\fi{}elj
  {et~al.}(2018)Gro\ifmmode~\check{s}\else \v{s}\fi{}elj, Mallet, Loureiro, \&
  Jenko}]{Groselj_al_2018}
Gro\ifmmode~\check{s}\else \v{s}\fi{}elj, D., Mallet, A., Loureiro, N.~F., \&
  Jenko, F. 2018, Phys. Rev. Lett., 120, 105101.
\newblock \url{https://link.aps.org/doi/10.1103/PhysRevLett.120.105101}

\bibitem[{Hadid {et~al.}(2018)Hadid, Sahraoui, Galtier, \&
  Huang}]{Hadid_al_2018}
Hadid, L.~Z., Sahraoui, F., Galtier, S., \& Huang, S.~Y. 2018, Phys. Rev.
  Lett., 120, 055102.
\newblock \url{https://link.aps.org/doi/10.1103/PhysRevLett.120.055102}

\bibitem[{{Hellinger} {et~al.}(2018){Hellinger}, {Verdini}, {Landi}, {Franci},
  \& {Matteini}}]{Hellinger_al_2018}
{Hellinger}, P., {Verdini}, A., {Landi}, S., {Franci}, L., \& {Matteini}, L.
  2018, \apjl, 857, L19

\bibitem[{{Karimabadi} {et~al.}(2013){Karimabadi}, {Roytershteyn}, {Wan},
  {Matthaeus}, {Daughton}, {Wu}, {Shay}, {Loring}, {Borovsky}, {Leon2ardis},
  {Chapman}, \& {Nakamura}}]{Karimabadi_al_2013}
{Karimabadi}, H., {Roytershteyn}, V., {Wan}, M., {et~al.} 2013, Phys.~Plasmas,
  20, doi:10.1063/1.4773205

\bibitem[{{Kiyani} {et~al.}(2009){Kiyani}, {Chapman}, {Khotyaintsev}, {Dunlop},
  \& {Sahraoui}}]{Kiyani_al_2009}
{Kiyani}, K.~H., {Chapman}, S.~C., {Khotyaintsev}, Y.~V., {Dunlop}, M.~W., \&
  {Sahraoui}, F. 2009, Phys.~Rev.~Lett., 103, arXiv:0906.2830

\bibitem[{{Kiyani} {et~al.}(2013){Kiyani}, {Chapman}, {Sahraoui}, {Hnat},
  {Fauvarque}, \& {Khotyaintsev}}]{Kiyani_al_2013}
{Kiyani}, K.~H., {Chapman}, S.~C., {Sahraoui}, F., {et~al.} 2013,
  Astrophys.~J., 763, arXiv:1008.0525

\bibitem[{{Kiyani} {et~al.}(2015){Kiyani}, {Osman}, \&
  {Chapman}}]{Kiyani_al_2015}
{Kiyani}, K.~H., {Osman}, K.~T., \& {Chapman}, S.~C. 2015, Philosophical
  Transactions of the Royal Society of London Series A, 373, 20140155

\bibitem[{Kobayashi {et~al.}(2017)Kobayashi, Sahraoui, Passot, Laveder, Sulem,
  Huang, Henri, \& Smets}]{Kobayashi_al_2017}
Kobayashi, S., Sahraoui, F., Passot, T., {et~al.} 2017, The Astrophysical
  Journal, 839, 122.
\newblock \url{https://doi.org/10.3847%2F1538-4357%2Faa67f2}

\bibitem[{{Le Contel} {et~al.}(2016){Le Contel}, {Leroy}, {Roux}, {Coillot},
  {Alison}, {Bouabdellah}, {Mirioni}, {Meslier}, {Galic}, {Vassal}, {Torbert},
  {Needell}, {Rau}, {Dors}, {Ergun}, {Westfall}, {Summers}, {Wallace},
  {Magnes}, {Valavanoglou}, {Olsson}, {Chutter}, {Macri}, {Myers}, {Turco},
  {Nolin}, {Bodet}, {Rowe}, {Tanguy}, \& {de la Porte}}]{LeContel_al_2016}
{Le Contel}, O., {Leroy}, P., {Roux}, A., {et~al.} 2016, Space Sci. Rev., 199,
  257

\bibitem[{Li {et~al.}(2016)Li, André, Khotyaintsev, Vaivads, Graham,
  Toledo-Redondo, Norgren, Henri, Wang, Tang, Lavraud, Vernisse, Turner, Burch,
  Torbert, Magnes, Russell, Blake, Mauk, Giles, Pollock, Fennell, Jaynes,
  Avanov, Dorelli, Gershman, Paterson, Saito, \& Strangeway}]{Li_al_2016}
Li, W., André, M., Khotyaintsev, Y.~V., {et~al.} 2016, Geophys.~Res.~Lett.,
  43, 5635.
\newblock
  \url{https://agupubs.onlinelibrary.wiley.com/doi/abs/10.1002/2016GL069192}

\bibitem[{{Lindqvist} {et~al.}(2016){Lindqvist}, {Olsson}, {Torbert}, {King},
  {Granoff}, {Rau}, {Needell}, {Turco}, {Dors}, {Beckman}, {Macri}, {Frost},
  {Salwen}, {Eriksson}, {{\AA}hl{\'e}n}, {Khotyaintsev}, {Porter},
  {Lappalainen}, {Ergun}, {Wermeer}, \& {Tucker}}]{Lindqvist_al_2016}
{Lindqvist}, P.-A., {Olsson}, G., {Torbert}, R.~B., {et~al.} 2016, Space Sci.
  Rev., 199, 137

\bibitem[{{MacBride} {et~al.}(2005){MacBride}, {Forman}, \&
  {Smith}}]{MacBride_al_2005}
{MacBride}, B.~T., {Forman}, M.~A., \& {Smith}, C.~W. 2005, in ESA Special
  Publication, Vol. 592, Solar Wind 11/SOHO 16, Connecting Sun and Heliosphere,
  ed. B.~{Fleck}, T.~H. {Zurbuchen}, \& H.~{Lacoste}, 613

\bibitem[{{Macek} {et~al.}(2018){Macek}, {Krasi{\'n}ska}, {Silveira}, {Sibeck},
  {Wawrzaszek}, {Burch}, \& {Russell}}]{Macek_al_2018}
{Macek}, W.~M., {Krasi{\'n}ska}, A., {Silveira}, M.~V.~D., {et~al.} 2018,
  \apjl, 864, L29

\bibitem[{{Matteini} {et~al.}(2017){Matteini}, {Alexandrova}, {Chen}, \&
  {Lacombe}}]{Matteini_al_2017}
{Matteini}, L., {Alexandrova}, O., {Chen}, C.~H.~K., \& {Lacombe}, C. 2017,
  mnras, 466, 945

\bibitem[{Matthaeus {et~al.}(2016)Matthaeus, Parashar, Wan, \&
  Wu}]{Matthaeus_al_2016}
Matthaeus, W.~H., Parashar, T.~N., Wan, M., \& Wu, P. 2016, The Astrophysical
  Journal, 827, L7.
\newblock \url{https://doi.org/10.3847%2F2041-8205%2F827%2F1%2Fl7}

\bibitem[{{Matthaeus} \& {Velli}(2011)}]{Matthaeus_Velli_2011}
{Matthaeus}, W.~H., \& {Velli}, M. 2011, Space Sci.~Rev., 160, 145

\bibitem[{{Matthaeus} {et~al.}(2015){Matthaeus}, {Wan}, {Servidio}, {Greco},
  {Osman}, {Oughton}, \& {Dmitruk}}]{Matthaeus_al_2015}
{Matthaeus}, W.~H., {Wan}, M., {Servidio}, S., {et~al.} 2015, Philosophical
  Transactions of the Royal Society of London Series A, 373, 20140154

\bibitem[{{Nakamura}(2019)}]{Nakamura_al_2019}
{Nakamura}, T.~K.~M. 2019, Solar/heliosphere 2, Magnetospheres of the solar
  system

\bibitem[{Nakamura {et~al.}(2013)Nakamura, Daughton, Karimabadi, \&
  Eriksson}]{Nakamura_al_2013}
Nakamura, T. K.~M., Daughton, W., Karimabadi, H., \& Eriksson, S. 2013, Journal
  of Geophysical Research: Space Physics, 118, 5742

\bibitem[{{Nakamura} {et~al.}(2017{\natexlab{a}}){Nakamura}, {Eriksson},
  {Hasegawa}, {Zenitani}, {Li}, {Genestreti}, {Nakamura}, \&
  {Daughton}}]{Nakamura_al_2017a}
{Nakamura}, T.~K.~M., {Eriksson}, S., {Hasegawa}, H., {et~al.}
  2017{\natexlab{a}}, J.~Geophys.~Res., 122, 11,505

\bibitem[{{Nakamura} {et~al.}(2017{\natexlab{b}}){Nakamura}, {Hasegawa},
  {Daughton}, {Eriksson}, {Li}, \& {Nakamura}}]{Nakamura_al_2017b}
{Nakamura}, T.~K.~M., {Hasegawa}, H., {Daughton}, W., {et~al.}
  2017{\natexlab{b}}, Nature Communications, 8, 1582

\bibitem[{Nakamura {et~al.}(2011)Nakamura, Hasegawa, Shinohara, \&
  Fujimoto}]{Nakamura_al_2011}
Nakamura, T. K.~M., Hasegawa, H., Shinohara, I., \& Fujimoto, M. 2011, Journal
  of Geophysical Research: Space Physics, 116, doi:10.1029/2010JA016046

\bibitem[{{Papini} {et~al.}(2019){Papini}, {Franci}, {Landi}, {Verdini},
  {Matteini}, \& {Hellinger}}]{Papini_al_2019}
{Papini}, E., {Franci}, L., {Landi}, S., {et~al.} 2019, \apj, 870, 52

\bibitem[{{Parashar} {et~al.}(2015){Parashar}, {Salem}, {Wicks}, {Karimabadi},
  {Gary}, \& {Matthaeus}}]{Parashar_al_2015}
{Parashar}, T.~N., {Salem}, C., {Wicks}, R.~T., {et~al.} 2015, Journal of
  Plasma Physics, 81, 905810513

\bibitem[{{Passot} {et~al.}(2015){Passot}, {Laveder}, \&
  {Sulem}}]{Passot_al_2015}
{Passot}, T., {Laveder}, D., \& {Sulem}, P.-L. 2015, Geophysical Reasearch
  Abstracts, 17, EGU2015

\bibitem[{{Passot} {et~al.}(2018){Passot}, {Sulem}, \&
  {Tassi}}]{Passot_al_2018}
{Passot}, T., {Sulem}, P.~L., \& {Tassi}, E. 2018, Phys.~Plasmas, 25, 042107

\bibitem[{{Phan} {et~al.}(2018){Phan}, {Eastwood}, {Shay}, {Drake}, {Sonnerup},
  {Fujimoto}, {Cassak}, {{\O}ieroset}, {Burch}, {Torbert}, {Rager}, {Dorelli},
  {Gershman}, {Pollock}, {Pyakurel}, {Haggerty}, {Khotyaintsev}, {Lavraud},
  {Saito}, {Oka}, {Ergun}, {Retino}, {Le Contel}, {Argall}, {Giles}, {Moore},
  {Wilder}, {Strangeway}, {Russell}, {Lindqvist}, \& {Magnes}}]{Phan_al_2018}
{Phan}, T.~D., {Eastwood}, J.~P., {Shay}, M.~A., {et~al.} 2018, nat, 557, 202

\bibitem[{{Podesta} {et~al.}(2007){Podesta}, {Roberts}, \&
  {Goldstein}}]{Podesta_al_2007}
{Podesta}, J.~J., {Roberts}, D.~A., \& {Goldstein}, M.~L. 2007, Astrophys.~J.,
  664, 543

\bibitem[{Politano \& Pouquet(1998)}]{Politano_Pouquet_1998}
Politano, H., \& Pouquet, A. 1998, Phys.~Rev.~E, 57, R21

\bibitem[{{Pollock} {et~al.}(2016){Pollock}, {Moore}, {Jacques}, {Burch},
  {Gliese}, {Saito}, {Omoto}, {Avanov}, {Barrie}, {Coffey}, {Dorelli},
  {Gershman}, {Giles}, {Rosnack}, {Salo}, {Yokota}, {Adrian}, {Aoustin},
  {Auletti}, {Aung}, {Bigio}, {Cao}, {Chandler}, {Chornay}, {Christian},
  {Clark}, {Collinson}, {Corris}, {De Los Santos}, {Devlin}, {Diaz},
  {Dickerson}, {Dickson}, {Diekmann}, {Diggs}, {Duncan}, {Figueroa-Vinas},
  {Firman}, {Freeman}, {Galassi}, {Garcia}, {Goodhart}, {Guererro}, {Hageman},
  {Hanley}, {Hemminger}, {Holland}, {Hutchins}, {James}, {Jones}, {Kreisler},
  {Kujawski}, {Lavu}, {Lobell}, {LeCompte}, {Lukemire}, {MacDonald}, {Mariano},
  {Mukai}, {Narayanan}, {Nguyan}, {Onizuka}, {Paterson}, {Persyn}, {Piepgrass},
  {Cheney}, {Rager}, {Raghuram}, {Ramil}, {Reichenthal}, {Rodriguez},
  {Rouzaud}, {Rucker}, {Saito}, {Samara}, {Sauvaud}, {Schuster}, {Shappirio},
  {Shelton}, {Sher}, {Smith}, {Smith}, {Smith}, {Steinfeld}, {Szymkiewicz},
  {Tanimoto}, {Taylor}, {Tucker}, {Tull}, {Uhl}, {Vloet}, {Walpole}, {Weidner},
  {White}, {Winkert}, {Yeh}, \& {Zeuch}}]{Pollock_al_2016}
{Pollock}, C., {Moore}, T., {Jacques}, A., {et~al.} 2016, Space Sci. Rev., 199,
  331

\bibitem[{{Pollock} {et~al.}(2018){Pollock}, {Burch}, {Chasapis}, {Giles},
  {Mackler}, {Matthaeus}, \& {Russell}}]{Pollock_al_2018}
{Pollock}, C.~J., {Burch}, J.~L., {Chasapis}, A., {et~al.} 2018, Journal of
  Atmospheric and Solar-Terrestrial Physics, 177, 84

\bibitem[{{Pucci} {et~al.}(2017){Pucci}, {Servidio}, {Sorriso-Valvo},
  {Olshevsky}, {Matthaeus}, {Malara}, {Goldman}, {Newman}, \&
  {Lapenta}}]{Pucci_al_2017}
{Pucci}, F., {Servidio}, S., {Sorriso-Valvo}, L., {et~al.} 2017, Astrophys.~J.,
  841, 60

\bibitem[{{Roberts} {et~al.}(2018){Roberts}, {Toledo-Redondo}, {Perrone},
  {Zhao}, {Narita}, {Gershman}, {Nakamura}, {Lavraud}, {Escoubet}, {Giles},
  {Dorelli}, {Pollock}, \& {Burch}}]{Roberts_al_2018}
{Roberts}, O.~W., {Toledo-Redondo}, S., {Perrone}, D., {et~al.} 2018, \grl, 45,
  7974

\bibitem[{Roytershteyn {et~al.}(2019)Roytershteyn, Boldyrev, Delzanno, Chen,
  Gro{\v{s}}elj, \& Loureiro}]{Roytershteyn_al_2019}
Roytershteyn, V., Boldyrev, S., Delzanno, G.~L., {et~al.} 2019, The
  Astrophysical Journal, 870, 103.
\newblock \url{https://doi.org/10.3847%2F1538-4357%2Faaf288}

\bibitem[{{Russell} {et~al.}(2016){Russell}, {Anderson}, {Baumjohann},
  {Bromund}, {Dearborn}, {Fischer}, {Le}, {Leinweber}, {Leneman}, {Magnes},
  {Means}, {Moldwin}, {Nakamura}, {Pierce}, {Plaschke}, {Rowe}, {Slavin},
  {Strangeway}, {Torbert}, {Hagen}, {Jernej}, {Valavanoglou}, \&
  {Richter}}]{Russell_al_2016}
{Russell}, C.~T., {Anderson}, B.~J., {Baumjohann}, W., {et~al.} 2016, Space
  Sci. Rev., 199, 189

\bibitem[{{Sahraoui} {et~al.}(2010){Sahraoui}, {Goldstein}, {Belmont}, {Canu},
  \& {Rezeau}}]{Sahraoui_al_2010}
{Sahraoui}, F., {Goldstein}, M.~L., {Belmont}, G., {Canu}, P., \& {Rezeau}, L.
  2010, Phys.~Rev.~Lett., 105, doi:10.1103/PhysRevLett.105.131101

\bibitem[{{Servidio} {et~al.}(2015){Servidio}, {Valentini}, {Perrone}, {Greco},
  {Califano}, {Matthaeus}, \& {Veltri}}]{Servidio_al_2015}
{Servidio}, S., {Valentini}, F., {Perrone}, D., {et~al.} 2015, J.~Plasma Phys.,
  81, 325810107

\bibitem[{Sharma~Pyakurel {et~al.}(2019)Sharma~Pyakurel, Shay, Phan, Matthaeus,
  Drake, TenBarge, Haggerty, Klein, Cassak, Parashar, Swisdak, \&
  Chasapis}]{SharmaPyakurel_al_2019}
Sharma~Pyakurel, P., Shay, M.~A., Phan, T.~D., {et~al.} 2019, Physics of
  Plasmas, 26, 082307

\bibitem[{{Sorriso-Valvo} {et~al.}(2007){Sorriso-Valvo}, {Marino}, {Carbone},
  {Noullez}, {Lepreti}, {Veltri}, {Bruno}, {Bavassano}, \&
  {Pietropaolo}}]{SorrisoValvo_al_2007}
{Sorriso-Valvo}, L., {Marino}, R., {Carbone}, V., {et~al.} 2007, Physical
  Review Letters, 99, 115001

\bibitem[{Sorriso-Valvo {et~al.}(2019)Sorriso-Valvo, Catapano, Retin\`o,
  Le~Contel, Perrone, Roberts, Coburn, Panebianco, Valentini, Perri, Greco,
  Malara, Carbone, Veltri, Pezzi, Fraternale, Di~Mare, Marino, Giles, Moore,
  Russell, Torbert, Burch, \& Khotyaintsev}]{SorrisoValvo_al_2019}
Sorriso-Valvo, L., Catapano, F., Retin\`o, A., {et~al.} 2019, Phys. Rev. Lett.,
  122, 035102.
\newblock \url{https://link.aps.org/doi/10.1103/PhysRevLett.122.035102}

\bibitem[{{Stawarz} {et~al.}(2016){Stawarz}, {Eriksson}, {Wilder}, {Ergun},
  {Schwartz}, {Pouquet}, {Burch}, {Giles}, {Khotyaintsev}, {Contel},
  {Lindqvist}, {Magnes}, {Pollock}, {Russell}, {Strangeway}, {Torbert},
  {Avanov}, {Dorelli}, {Eastwood}, {Gershman}, {Goodrich}, {Malaspina},
  {Marklund}, {Mirioni}, \& {Sturner}}]{Stawarz_al_2016}
{Stawarz}, J.~E., {Eriksson}, S., {Wilder}, F.~D., {et~al.} 2016,
  J.~Geophys.~Res., 121, 11

\bibitem[{{Stawarz} {et~al.}(2019){Stawarz}, {Eastwood}, {Phan}, {Gingell},
  {Shay}, {Burch}, {Ergun}, {Giles}, {Gershman}, {Le Contel}, {Lindqvist},
  {Russell}, {Strangeway}, {Torbert}, {Argall}, {Fischer}, {Magnes}, \&
  {Franci}}]{Stawarz_al_2019}
{Stawarz}, J.~E., {Eastwood}, J.~P., {Phan}, T.~D., {et~al.} 2019,
  Astrophys.~J.~Lett., 877, L37

\bibitem[{{Sturner} {et~al.}(2018){Sturner}, {Eriksson}, {Nakamura},
  {Gershman}, {Plaschke}, {Ergun}, {Wilder}, {Giles}, {Pollock}, {Paterson},
  {Strangeway}, {Baumjohann}, \& {Burch}}]{Sturner_al_2018}
{Sturner}, A.~P., {Eriksson}, S., {Nakamura}, T., {et~al.} 2018,
  J.~Geophys.~Res., 123, 1305

\bibitem[{{Taylor}(1938)}]{Taylor_al_1938}
{Taylor}, G.~I. 1938, Proceedings of the Royal Society of London Series A, 164,
  476

\bibitem[{{TenBarge} \& {Howes}(2013)}]{TenBarge_Howes_2013}
{TenBarge}, J.~M., \& {Howes}, G.~G. 2013, Astrophys.~J.~Lett., 771, L27

\bibitem[{Told {et~al.}(2015)Told, Jenko, TenBarge, Howes, \&
  Hammett}]{Told_al_2015}
Told, D., Jenko, F., TenBarge, J.~M., Howes, G.~G., \& Hammett, G.~W. 2015,
  Phys. Rev. Lett., 115, 025003.
\newblock \url{https://link.aps.org/doi/10.1103/PhysRevLett.115.025003}

\bibitem[{{{\v S}afr{\'a}nkov{\'a}} {et~al.}(2016){{\v S}afr{\'a}nkov{\'a}},
  {N{\v e}me{\v c}ek}, {N{\v e}mec}, {P{\v r}ech}, {Chen}, \&
  {Zastenker}}]{Safrankova_al_2016}
{{\v S}afr{\'a}nkov{\'a}}, J., {N{\v e}me{\v c}ek}, Z., {N{\v e}mec}, F.,
  {et~al.} 2016, Astrophys.~J., 825, 121

\bibitem[{{{\v S}afr{\'a}nkov{\'a}} {et~al.}(2015){{\v S}afr{\'a}nkov{\'a}},
  {N{\v e}me{\v c}ek}, {N{\v e}mec}, {P{\v r}ech}, {Pit{\v n}a}, {Chen}, \&
  {Zastenker}}]{Safrankova_al_2015}
---. 2015, Astrophys.~J., 803, 107

\bibitem[{{Valentini, F.} {et~al.}(2017){Valentini, F.}, {V\'asconez, C. L.},
  {Pezzi, O.}, {Servidio, S.}, {Malara, F.}, \& {Pucci,
  F.}}]{Valentini_al_2017}
{Valentini, F.}, {V\'asconez, C. L.}, {Pezzi, O.}, {et~al.} 2017, A\&A, 599,
  A8.
\newblock \url{https://doi.org/10.1051/0004-6361/201629240}

\bibitem[{V\'asconez {et~al.}(2014)V\'asconez, Valentini, Camporeale, \&
  Veltri}]{Vasconez_al_2014}
V\'asconez, C.~L., Valentini, F., Camporeale, E., \& Veltri, P. 2014, Physics
  of Plasmas, 21, 112107

\bibitem[{{Vasquez} {et~al.}(2014){Vasquez}, {Markovskii}, \&
  {Chandran}}]{Vasquez_al_2014}
{Vasquez}, B.~J., {Markovskii}, S.~A., \& {Chandran}, B.~D.~G. 2014,
  Astrophys.~J., 788, 178

\bibitem[{Verdini {et~al.}(2015)Verdini, Grappin, Hellinger, Landi, \&
  {M\"uller}}]{Verdini_al_2015}
Verdini, A., Grappin, R., Hellinger, P., Landi, S., \& {M\"uller}, W.~C. 2015,
  Astrophys.~J., 804, doi:10.1088/0004-637X/804/2/119

\bibitem[{{Vernisse} {et~al.}(2016){Vernisse}, {Lavraud}, {Eriksson},
  {Gershman}, {Dorelli}, {Pollock}, {Giles}, {Aunai}, {Avanov}, {Burch},
  {Chandler}, {Coffey}, {Dargent}, {Ergun}, {Farrugia}, {G{\'e}not}, {Graham},
  {Hasegawa}, {Jacquey}, {Kacem}, {Khotyaintsev}, {Li}, {Magnes}, {Marchaudon},
  {Moore}, {Paterson}, {Penou}, {Phan}, {Retino}, {Russell}, {Saito},
  {Sauvaud}, {Torbert}, {Wilder}, \& {Yokota}}]{Vernisse_al_2016}
{Vernisse}, Y., {Lavraud}, B., {Eriksson}, S., {et~al.} 2016, J.~Geophys.~Res.,
  121, 9926

\bibitem[{{Verscharen} {et~al.}(2019){Verscharen}, {Klein}, \&
  {Maruca}}]{Verscharen_al_2019}
{Verscharen}, D., {Klein}, K.~G., \& {Maruca}, B.~A. 2019, arXiv e-prints,
  arXiv:1902.03448

\bibitem[{V{\"o}r{\"o}s {et~al.}(2006)V{\"o}r{\"o}s, Baumjohann, Nakamura,
  Volwerk, \& Runov}]{Voros_al_2006}
V{\"o}r{\"o}s, Z., Baumjohann, W., Nakamura, R., Volwerk, M., \& Runov, A.
  2006, Space Science Reviews, 122, 301.
\newblock \url{https://doi.org/10.1007/s11214-006-6987-7}

\bibitem[{{Wan} {et~al.}(2015){Wan}, {Matthaeus}, {Roytershteyn}, {Karimabadi},
  {Parashar}, {Wu}, \& {Shay}}]{Wan_al_2015}
{Wan}, M., {Matthaeus}, W.~H., {Roytershteyn}, V., {et~al.} 2015, Physical
  Review Letters, 114, 175002

\bibitem[{{Wilder} {et~al.}(2016){Wilder}, {Ergun}, {Schwartz}, {Newman},
  {Eriksson}, {Stawarz}, {Goldman}, {Goodrich}, {Gershman}, {Malaspina},
  {Holmes}, {Sturner}, {Burch}, {Torbert}, {Lindqvist}, {Marklund},
  {Khotyaintsev}, {Strangeway}, {Russell}, {Pollock}, {Giles}, {Dorrelli},
  {Avanov}, {Patterson}, {Plaschke}, \& {Magnes}}]{Wilder_al_2016}
{Wilder}, F.~D., {Ergun}, R.~E., {Schwartz}, S.~J., {et~al.} 2016,
  Geophys.~Res.~Lett., 43, 8859

\bibitem[{{Wu} {et~al.}(2013){Wu}, {Perri}, {Osman}, {Wan}, {Matthaeus},
  {Shay}, {Goldstein}, {Karimabadi}, \& {Chapman}}]{Wu_al_2013}
{Wu}, P., {Perri}, S., {Osman}, K., {et~al.} 2013, Astrophys. J. Let., 763,
  doi:10.1088/2041-8205/763/2/L30

\bibitem[{Zhdankin {et~al.}(2019)Zhdankin, Uzdensky, Werner, \&
  Begelman}]{Zhdankin_al_2019}
Zhdankin, V., Uzdensky, D.~A., Werner, G.~R., \& Begelman, M.~C. 2019, Phys.
  Rev. Lett., 122, 055101.
\newblock \url{https://link.aps.org/doi/10.1103/PhysRevLett.122.055101}

\bibitem[{Zimbardo {et~al.}(2010)Zimbardo, Greco, Sorriso-Valvo, Perri,
  V{\"o}r{\"o}s, Aburjania, Chargazia, \& Alexandrova}]{Zimbardo_al_2010}
Zimbardo, G., Greco, A., Sorriso-Valvo, L., {et~al.} 2010, Space Science
  Reviews, 156, 89.
\newblock \url{https://doi.org/10.1007/s11214-010-9692-5}

\end{thebibliography}



\end{document}